\documentclass[12pt,preprint]{aastex}

\usepackage{epsf}

\def\deg{\ifmmode {^\circ}\else {$^\circ$}\fi}
\def\degree{\ifmmode {^\circ}\else {$^\circ$}\fi}
\def\mum{\ifmmode {\rm \,\mu {\rm m}}\else $\rm \,\mu {\rm m}$\fi}
\def\micron{\ifmmode {\rm \,\mu {\rm m}}\else $\rm \,\mu {\rm m}$\fi}
\def\arcsec{\ifmmode ^{\prime \prime}\else $^{\prime \prime}$\fi}

\def\inch{\ifmmode ^{\prime \prime}\else $^{\prime \prime}$\fi}
\def\Msun{\ifmmode {M_{\odot}}\else $M_{\odot}$\fi}
\def\Mstar{\ifmmode {M_{*}}\else $M_{*}$\fi}
\def\lsun{\ifmmode {\rm L_{\odot}}\else $\rm L_{\odot}$\fi}
\def\mstar{\ifmmode {\rm M_{\star}}\else $\rm M_{\star}$\fi}
\def\lstar{\ifmmode {\rm L_{\star}}\else $\rm L_{\star}$\fi}
\def\md{\ifmmode {\rm M_d}\else $\rm M_d$\fi}
\def\ld{\ifmmode {\rm L_d}\else $\rm L_d$\fi}
\def\mearth{\ifmmode {\rm M_{\oplus}}\else $\rm M_{\oplus}$\fi}
\def\qdstar{\ifmmode Q_D^\star\else $Q_D^\star$\fi}
\def\kms{\ifmmode {\rm \,km\,s^{-1}}\else $\rm \,km\,~s^{-1}$\fi}
\def\ms{\ifmmode {\rm m~s^{-1}}\else $\rm m~s^{-1}$\fi}
\def\mesc{\ifmmode m_{esc}\else $m_{esc}$\fi}
\def\rmin{\ifmmode r_{min}\else $r_{min}$\fi}
\def\rmax{\ifmmode r_{max}\else $r_{max}$\fi}
\def\mmin{\ifmmode m_{min}\else $m_{min}$\fi}
\def\mmax{\ifmmode m_{max}\else $m_{max}$\fi}
\def\rmind{\ifmmode r_{min,d}\else $r_{min,d}$\fi}
\def\rmaxd{\ifmmode r_{max,d}\else $r_{max,d}$\fi}
\def\mmaxd{\ifmmode m_{max,d}\else $m_{max,d}$\fi}
\def\vrad{\ifmmode v_{rad}\else $v_{rad}$\fi}
\def\qz{\ifmmode q_{0}\else $q_{0}$\fi}
\def\qi{\ifmmode q_{i}\else $q_{i}$\fi}
\def\ql{\ifmmode q_{l}\else $q_{l}$\fi}
\def\qs{\ifmmode q_{s}\else $q_{s}$\fi}
\def\rbrk{\ifmmode r_{brk}\else $r_{brk}$\fi}
\def\rdamp{\ifmmode r_{damp}\else $r_{damp}$\fi}
\def\ain{\ifmmode a_{in}\else $a_{in}$\fi}
\def\aout{\ifmmode a_{out}\else $a_{out}$\fi}
\def\r0{\ifmmode r_{0}\else $r_{0}$\fi}
\def\m0{\ifmmode m_{0}\else $m_{0}$\fi}
\def\M0{\ifmmode M_{0}\else $M_{0}$\fi}
\def\xm{\ifmmode x_{m}\else $x_{m}$\fi}
\def\gyr{\ifmmode {\rm g~yr^{-1}}\else ${\rm g~yr^{-1}}$\fi}
\def\cms{\ifmmode {\rm cm~s^{-1}}\else ${\rm cm~s^{-1}}$\fi}
\def\gcms{\ifmmode {\rm g~cm^{-2}}\else $\rm g~cm^{-2}$\fi}
\def\gcmc{\ifmmode {\rm g~cm^{-3}}\else $\rm g~cm^{-3}$\fi}
\def\pcm{\ifmmode {\rm \,cm^{-1}}\else $\rm \,cm^{-1}$\fi}
\def\psqcm{\ifmmode {\rm \,cm^{-2}}\else $\rm \,cm^{-2}$\fi}
\def\pccm{\ifmmode {\rm \,cm^{-3}}\else $\rm \,cm^{-3}$\fi}
\def\water{\ifmmode {\rm H_2O}\else $\rm H_2O$\fi}
\def\hm{\ifmmode {\rm H_2}\else $\rm H_2$\fi}
\def\ctwohtwo{\ifmmode {\rm C_2H_2}\else $\rm C_2H_2$\fi}
\def\cotwo{\ifmmode {\rm CO_2}\else $\rm CO_2$\fi}
\def\Lya{\ifmmode {\rm Ly\alpha}\else Ly$\alpha$\fi}
\def\as205n{AS\,205\,N}

\shorttitle{}
\shortauthors{}

\begin{document}


\title{Spectrally Resolved Mid-Infrared Molecular Emission 
from Protoplanetary Disks and the Chemical Fingerprint of 
Planetesimal Formation}


\author{Joan R.\ Najita}
\affil{National Optical Astronomy Observatory, 950 N.\ Cherry Avenue,
Tucson, AZ 85719, USA}

\author{John S.\ Carr}
\affil{Naval Research Laboratory, Code 7213, Washington, DC 20375, USA}

\author{Colette Salyk}
\affil{Vassar College, 124 Raymond Avenue, Poughkeepsie, NY 12604, USA} 

\author{John H.\ Lacy}
\affil{Department of Astronomy, University of Texas at Austin, Austin, TX 78712, USA}

\author{Matthew J. Richter}
\affil{Physics Department, University of California at Davis, Davis, CA 95616, USA}

\and

\author{Curtis DeWitt}
\affil{USRA/SOFIA, NASA Ames Research Center, MS 232-12, Bldg.\ 232, Rm.\ 130-31, Moffett Field, CA 94035-000}


\begin{abstract}
We present high resolution spectroscopy of 
mid-infrared molecular emission 
from two very active T Tauri stars, \as205n\ and DR Tau. 
In addition to 
measuring high signal-to-noise line profiles of water, 
we report the first spectrally resolved mid-infrared 
line profiles of HCN emission from protoplanetary disks.  
The similar line profiles and temperatures of the 
HCN and water emission indicate that they arise in the same 
volume of the disk atmosphere, within 1--2\,AU of the star.  
The results support the earlier suggestion that 
the observed trend of increasing HCN/water emission 
with disk mass is a chemical fingerprint of 
planetesimal formation and core accretion in action. 
In addition to directly constraining the emitting 
radii of the molecules, the high resolution spectra also 
help to break degeneracies between temperature and 
column density in deriving molecular abundances from 
low resolution mid-infrared spectra. 
As a result, they can improve our understanding 
of the extent to which inner disks are 
chemically active.
Contrary to predictions from HCN excitation studies 
carried out for \as205n, 
the mid-infrared and near-infrared line profiles of HCN 
are remarkably similar. 
The discrepancy may indicate that HCN is not abundant beyond a 
couple of AU or that infrared pumping of HCN does not dominate 
at these distances.

\end{abstract}





\section{Introduction}

Mid-infrared emission from water and simple organic molecules 
(HCN, \ctwohtwo) is commonly detected in 
spectra of classical T Tauri stars (CTTS) 
measured by the {\it Spitzer Space Telescope.} 
The high critical densities of the detected lines ($>10^9 \pccm$), 
the warm temperature of the emitting gas (300--1000\,K), and 
its modest inferred emitting area
argue that the emission arises from the inner few AU of the disk 
(Carr \& Najita 2011; Salyk et al.\ 2011a; Pontoppidan et al.\ 2010b), 
the region within the snowline. 
These observations complement 
not only 
near-infrared rovibrational observations of 
CO (4.5\micron; Banzatti \& Pontoppidan 2015; J.\ Brown et al.\ 2013; 
Salyk et al.\ 2011b; Bast et al.\ 2011; Pontoppidan et al.\ 2011; 
Najita et al.\ 2003), 
water and OH (3\micron; 
Salyk et al.\ 2008; 
Mandell et al.\ 2012; 
Doppmann et al.\ 2011; 
L.\ Brown et al.\ 2013),  
and organic molecules (3\micron; 
Mandell et al.\ 2012; 
Gibb \& Horne 2013; 
Doppmann et al.\ 2008; 
Gibb et al.\ 2007), 
which are generally sensitive to 
warmer gas at smaller disk radii, but also   
far-infrared and submillimeter observations that probe 
disks beyond the snowline (e.g., \"Oberg et al.\ 2015). 

The properties of the spectrally unresolved {\it Spitzer} spectra have
been used to probe the chemical state of inner disks  
and their planet formation status. 
In some analyses, 
the abundance ratios of the organics relative to water,  
as inferred from simple slab models of the emission,  
are enhanced compared to the molecular abundances of comets
(Carr \& Najita 2008, 2011),  
which serve as surrogate probes 
of the more distant giant planet region of the disk, 
beyond the snowline.   
The enhanced abundance ratios of inner disks 
are interpreted as evidence for an active inner disk chemistry 
(e.g., Carr \& Najita 2008).
That is, disks synthesize molecules in their warm, high density 
inner regions rather than merely inheriting their organic inventory 
from larger disk radii or from molecular clouds 
(e.g., Pontoppidan et al.\ 2014). 

The {\it Spitzer} molecular emission properties may also 
encode chemical evidence for the formation of icy planetesimals, 
a potential signature of core accretion in action. 
As the building blocks of planets, 
planetesimals are fundamental to 
the core accretion picture of planet formation,
but they are observationally elusive: 
it is difficult to detect a kilometer-sized rock or even a 
Mars-sized protoplanet embedded in a disk! 
These bodies are neither self-luminous nor large enough to 
detect directly. 
They are also not massive enough to produce an observable 
dynamical signature, e.g., by opening a gap in the disk. 

The {\it Spitzer} molecular emission properties show an 
interesting trend in this regard.  
The ratio of HCN/\water\ emission strength 
increases with disk mass, a trend that has been
interpreted as a possible chemical fingerprint of planetesimal formation
(Najita et al.\ 2013). Because higher mass disks are expected to 
form icy planetesimals and protoplanets more readily 
in the giant planet region of the disk, 
and thereby sequester water and oxygen as ice beyond the snowline, 
the accreting oxygen-poor disk gas leads to a 
carbon-rich gaseous inner disk. 
Modest variations in the C/O ratio of the inner disk can, 
in principle, induce large variations in the molecular abundances 
of the inner disk atmosphere at a level that is 
consistent with the range of observed molecular emission flux ratios 
(Najita et al.\ 2011).  

The above inferences assume that the mid-infrared organics and 
water emission probe the same region of the disk 
(i.e., the same disk radii and vertical height in the disk).  
In reality,
the spectrally and spatially unresolved {\it Spitzer} data  
can potentially constrain only molecular emitting areas 
(Carr \& Najita 2011), and emission radii are unconstrained. 
The latter 
can be inferred more directly from spectrally resolved emission 
line profiles.  
Here we test the assumption from Najita et al.\ (2013) 
that the organics and water emission arise from the same region 
by spectrally resolving the mid-infrared HCN and \water\ 
line emission from two CTTS.

Previous high spectral resolution studies of inner disk atmospheres 
have spectrally resolved bright water emission in a few sources,  
finding that the mid-infrared line profiles 
are consistent with emission from gas in Keplerian rotation over 
a range of radii consistent with the emitting area inferred from  
simple slab models 
(Knez et al.\ 2007;  
Pontoppidan et al.\ 2010b; 
Salyk et al.\ 2015; 
Carr et al.\ 2018, in preparation).
The present study complements these studies of water emission 
and resolves the line profiles of the much fainter HCN emission. 

We describe our observations in \S 2 
and our results in \S 3. 
Section 4 discusses 
how the results bear on 
our ability to probe 
the chemical fingerprint of planetesimal formation 
(\S 4.3), 
to measure the abundance ratios of inner disks (\S 4.4), 
and 
our understanding of 
the excitation of the infrared transitions of HCN (\S 4.5). 
We conclude with a summary (\S 5).

\section{Observations}

DR Tau and \as205n\ are very active CTTS with bright molecular emission,
as observed with {\it Spitzer} and ground-based spectroscopy (Salyk et
al.\ 2008, Mandell et al.\ 2012; Pontoppidan et al.\ 2010ab; Banzatti
et al.\ 2014; J.\ Brown et al.\ 2013). \as205n\ has also been well
studied at millimeter wavelengths (e.g., Andrews et al.\ 2009; Salyk et
al.\ 2014).

The detection with {\it Spitzer} of MIR molecular emission from 
\as205n\ and DR Tau, reported initially by Salyk et al.\ (2008),  
has been studied in greater detail in subsequent analyses of 
{\it Spitzer} spectra (Pontoppidan et al.\ 2010a, Salyk et al.\ 2011a). 
The water emission from \as205n\ has been studied previously at high 
spectral resolution using the VLT:  
in the $L$-band with CRIRES ($R\sim$100,000) 
and in the MIR with VISIR ($R\sim$20,000; Pontoppidan et al.\ 2010b).  
MIR water emission from DR Tau has been studied previously 
with VISIR (Banzatti et al.\ 2014). 
Mandell et al.\ (2012) studied the  
$L$-band water and organics emission 
from \as205n\ (CRIRES; $R\approx$96,000) 
and DR Tau (Keck/NIRSPEC; $R\approx$25,000) 
at lower signal-to-noise than in the present study. 

We observed \as205n\ and DR Tau using the high-resolution ($R\sim$80,000), 
cross-dispersed
mode of the Texas Echelon Cross Echelle Spectrograph (TEXES; Lacy
et al.\ 2002) on the Gemini-North 8m telescope (Table 1).
For all observations, we used a 0.5\arcsec\ wide slit and nodded along the
slit to remove background emission.  We followed the standard TEXES
procedure of observing a blackbody and blank sky roughly every 6
minutes to provide wavelength calibration, flat fielding, 
and approximate flux calibration.
In addition, we observed asteroids as telluric
standards for each object.  The asteroid observations also improve
the removal of the instrumental signature, as they are point sources,
whereas the blackbody is an extended object.

DR Tau was observed on 17 and 19 November 2013 (UT) during the TEXES
visitor-instrument campaign at Gemini.  We set the central wavenumber
of the instrument to roughly 781\pcm\ on the first night to detect
the HCN R(23) line at 782.653\pcm\ and to  roughly 805\pcm\ on the
second night to detect several \water\ lines.  At these wavelengths,
the spectral orders are wider than the TEXES detector, so there
are small gaps in the spectral coverage.  
Similarly, we observed \as205n\ on 17 and 18 August 2014 (UT) during
the next TEXES visitor-instrument campaign at Gemini.  The spectral
settings were similar to but slightly different from those for DR Tau,  
with the HCN R(22) line included in the 781\pcm\ setting. 

\begin{deluxetable}{llllll}
\tabletypesize{\footnotesize}
\tablecaption{\label{t:std}Observation Log}
\tablehead{
Target & Date & $\lambda$ Setting & $t_{\rm Int}$ & Telluric Standard & Features Targeted}
\startdata 
DR Tau	& 17 Nov 2013 & 781\pcm\ & 41 min & 10 Hygiea & HCN R(23), \ctwohtwo\ R(21) \\
DR Tau	& 19 Nov 2013 & 805\pcm\ & 20.5 min & 10 Hygiea & \water\ \\
\as205n\	& 17 Aug 2014 & 781\pcm\ & 55.\ min& 15 Eunomia, 16 Psyche & HCN R(22), R(23), \ctwohtwo\ R(21)\\
\as205n\	& 18 Aug 2014 & 805\pcm\ & 17.3 min & 15 Eunomia & \water\ \\
\as205n\	& 18 Aug 2014 & 781\pcm\ & 35.6 min & 16 Psyche & HCN R(22), R(23), \ctwohtwo\ R(21) \\
\enddata
\end{deluxetable}

We reduced the data using the standard TEXES pipeline (Lacy et al.\
2002).  The pipeline corrects for spikes, differences nod pairs,
aligns the stellar continuum along the slit in each nod difference,
sums the nod differences, and extracts the spectrum after weighting
by the distribution of continuum emission along the slit.  In
addition, the user  interacts with the pipeline to set the frequency
scale based on atmospheric features. 
The frequency scale is accurate throughout the setting to 1\,\kms.

Before dividing the stellar spectrum by that of the telluric standard, 
the asteroid spectra were raised to a power in order to account 
for differences in the airmass and precipitable water vapor 
at which the target and calibrator were observed. 
This scaling 
makes sense in a simple plane parallel 
atmosphere, where the atmospheric transmission 
at zenith distance $z$ 
is related to the transmission at zenith, $e^{-\tau_0}$, 
by $e^{-\tau(z)}  
= (e^{-\tau_0})^{\sec z}$.
Differences in airmass and water vapor column affect the strength of 
telluric atmospheric lines.

The DR Tau spectrum was flux calibrated by scaling the continuum
to 1.87\,Jy, based on the {\it Spitzer} IRS SH spectrum. 
Studies of the spectral variability of DR Tau
show that the flux varies by $\sim 20$\% at this wavelength 
(Kospal et al.\ 2012; Banzatti et al.\ 2014).
For \as205n, we scaled the continuum to 6.0\,Jy.
This value was based on the IRS SH spectrum (\S 3.2) and the observed
history of the primary/secondary flux ratio at 12.5\micron\
(Liu et al.\ 1996; McCabe et al.\ 2006). The {\it Spitzer} 
IRS flux is
consistent with most published values for the combined flux of
the binary components, although it has been observed to be brighter
by 60\% (Liu et al.\ 1996).
Additional details are provided in \S 3.2.

\section{Results}

Table 2 reports the detected spectral features.
In the 781\pcm\ setting on \as205n, we detected 
the R22 and R23 lines of the HCN $\nu_2$ band, 
the R21 line of the \ctwohtwo\ $\nu_5$ band, as well as  
rotational water lines at 779.304\pcm\ and 783.762\pcm\ 
(Fig.~1; hereafter 779\pcm\ and 784\pcm). 
The 784\pcm\ water line is truncated by the edge of the order, 
and the 779\pcm\ line is near the edge of the order and is 
affected by telluric absorption.
In the 805\pcm\ setting, we detected the rotational water lines at 
808.083\pcm, 806.696\pcm, and 805.994\pcm\ at high signal-to-noise 
(hereafter 808\pcm, 807\pcm, and 806\pcm)
as well as lower energy water lines  at 
803.546\pcm\ and 802.990\pcm\ 
(Fig.~2; hereafter 804\pcm\ and 803\pcm). 
The latter two lines have higher noise as a result of 
telluric absorption.

In the 781\pcm\ setting on DR Tau, we detected 
the HCN R23 line  
but not the water lines at 779\pcm\ and 784\pcm\ 
(Fig.~3).  
We were unable to study the HCN R22 line, 
which fell in a gap between two orders. 
The \ctwohtwo\ R21 line was not detected.
This is not surprising given the far lower 
signal-to-noise of the HCN profile
and the smaller \ctwohtwo/HCN flux ratio at 
low spectral resolution of DR Tau compared to \as205n.
In their {\it Spitzer} IRS spectra, 
the \ctwohtwo/HCN Q band flux ratios are $\sim 0.5$ and 
$\sim 1$ for DR Tau and \as205n,
respectively (Salyk et al.\ 2011a).
In the 805\pcm\ setting, we detected 
the water lines at 
808\pcm, 807\pcm, and 806\pcm\ (Fig.~4). 
Compared to the \as205n\ spectrum, 
the 803\pcm\ and 804\pcm\ water lines 
in the DR Tau spectrum 
are more affected by telluric absorption 
because of the smaller velocity shift of DR Tau at the 
epoch of observation ($\sim 17\kms$) compared to \as205n\ 
($\sim 25 \kms$).
As a result, the 803\pcm\ line is
detected in the DR Tau spectrum, 
but the 804\pcm\ line is lost in the telluric absorption.

\begin{deluxetable}{lcllll}
\tabletypesize{\footnotesize}
\tablecaption{\label{t:std}Detected Features}
\tablehead{
Line & Rest Waveno & $E_{\rm up}$ & $A_{ul}$ & \as205n\ & DR Tau \\
     & (cm$^{-1}$) & (K)          & s$^{-1}$ & 	   & }
\startdata 
{\bf 781\pcm\ Setting:}    & & & \\ 
HCN R22 v$_2$=1-0 	   & 779.727 	& 2197	& 1.36	& e		& in gap 	\\
\ctwohtwo\ R21 v$_5$=1-0   & 780.753 	& 1905	& 3.82	& e		& x	\\
HCN R23 v$_2$=1-0 	   & 782.653 	& 2299	& 1.38	& e		& e	\\
\ctwohtwo\ R22 v$_5$=1-0   & 783.087 	& 1983	& 3.86	& in gap	& on edge	\\
\water\ 10,8,2-9,5,5 (p)   & 779.304 	& 3244	& 0.16	& e		& poor correction	\\
\water\ 17,6,12-16,3,13 (p)& 783.762	& 6073	& 9.92	& on edge	& x	\\
& & & \\
{\bf 805\pcm\ Setting:}    & & & \\
\water\ 13,7,6-12,4,9 (o)  & 802.990   	& 4213	& 1.05	& e	 & e? poor correction 	\\
\water\ 11,8,3-10,5,6 (o)  & 803.546	& 3629	& 0.29	& e	 & e? poor correction 	\\
\water\ 16,3,13-15,2,14 (p)& 805.994 	& 4946	& 4.22	& e	 & e 	\\
\water\ 17,4,13-16,3,14 (o)& 806.696   	& 5781	& 7.67	& e	 & e 	\\
\water\ 16,4,13-15,1,14	(o)& 808.038 	& 4949	& 4.21	& e	 & e 	\\
\enddata
\tablecomments{Rest wavenumbers are from HITRAN (Rothman et al.\ 2013). 
In columns (4) and (5), `e' indicates emission, `x' non-detection, 
and `in gap' 
and `on edge' indicate lines that fell either in a gap between two orders or 
on the edge of an order.}
\end{deluxetable}

\subsection{Line Profiles and Velocities}

To determine the HCN and \water\ emission line profiles, 
we first subtracted a local linear slope to the continuum, 
determined from neighboring wavelength 
regions located just beyond the line emission region.
The profiles of the bright \water\ lines in the 805\pcm\ setting 
were used to set the velocity extent of the line emission region 
($\pm\, 25\kms$).

The line profiles of the water and organic emission in 
both the \as205n\ and DR Tau 
spectra are consistent with each other within the noise. 
As shown in Figure 5,
the HCN line profile in the \as205n\ spectrum 
is consistent with that of the \ctwohtwo\ emission 
and with that of the weak \water\ line at 779\pcm. 
Because the profiles of the HCN R22 and R23 lines for \as205n\ 
were similar, we averaged them together to obtain 
a higher signal-to-noise profile; 
the individual lines were scaled to a common line flux and the 
profiles averaged. 
The HCN emission has a FWHM of $\sim 20\,\kms$ and 
peaks at 
$v_{\rm helio} \sim -4\,\kms.$ 
The top panel of Figure 5 compares the average line profile 
of the HCN lines (R22 and R23) with 
the profile of the \ctwohtwo\ R21 line. 
The lower panel of Figure 5 compares the average of 
all three organic lines (HCN R22, HCN R23, and \ctwohtwo\ R21) 
with the 779\pcm\ water line. 
The \water\ 779\pcm\ and HCN lines 
are similar in strength and
detected at a similar signal-to-noise ratio.

The bright water lines in the 805\pcm\ setting on \as205n\
provide a better estimate of the water line profile. 
To define the water line profile, 
we examined the higher energy lines, which have good telluric correction 
(808\pcm, 807\pcm, and 806\pcm), 
and excluded the lower energy water lines (803\pcm, 804\pcm), 
which have stronger telluric absorption and less certain line profiles. 
Figure 6 compares the average HCN line profile 
with the profile of the 808\pcm\ \water\ line,  
which is closer in excitation to the HCN lines.  
The profiles of the two other water lines compare similarly. 
All three water lines are discussed in greater detail  
in J.\ S.\ Carr et al.\ (2018, in preparation).

Given the wavelength calibration uncertainty and the 
limited signal-to-noise ratio of the HCN profile, the water and HCN 
profiles appear consistent with each other.  
Echoing the result found here, Mandell et al.\ (2012) found 
that line profiles of the HCN and \water\ emission lines 
observed in the $L$-band with CRIRES were consistent with each 
other, although at a much lower signal-to-noise ratio.

Similar results are found for the water and HCN emission 
from DR Tau. 
Figure 7 shows the average water line profile, obtained by 
scaling the three bright water lines 
in the 805\pcm\ setting 
to the same equivalent width and averaging.  
The resulting profile   
is centrally peaked, 
with a FWHM of $13\,\kms$ and centered at $v_{\rm helio}$ of $25\,\kms$. 
The HCN line profile  
overlaps the \water\ lines in velocity and  
appears consistent with the \water\ profile
given the limited signal-to-noise of the detection.

The water line emission properties we measure are in good agreement 
with earlier observations of water emission from DR Tau. 
Observations with VISIR found the same $v_{\rm helio}$ 
for the 806\pcm\ and 805\pcm\ water lines 
(Banzatti et al.\ 2014). 
Our measured FWHM of the MIR water emission ($13\kms$) 
is consistent with 
that of the $L$-band water emission from DR Tau 
measured with Keck/NIRSPEC ($18\kms$; L.\ Brown et al.\ 2013) 
given the lower spectral resolution of the NIRSPEC spectrum 
($12\kms$). 
From the NIRSPEC spectrum, Brown et al.\ (2013)
inferred that the water emission 
extends outward in disk radius to at least 
0.4\,AU, i.e., to a projected velocity of at least $7\kms$ for the 
stellar mass ($0.4\Msun$; Isella et al.\ 2009) 
and source inclination ($i=13^\circ$; see also Pontoppidan et al.\ 2011). 
At the higher resolution of the TEXES observations, we find that 
the MIR water emission extends to lower velocities and 
therefore larger disk radii (\S 4.1).

More generally, the line center velocities of the molecular emission
from DR Tau and \as205n\ are consistent, within the TEXES wavelength
calibration uncertainty, with that of other IR and submillimeter
molecular emission from the sources and with their stellar velocities.
Further details are provided in Appendix A.

\subsection{Characterizing the Emission with Simple Slab Models}

To characterize the detected emission, we compared the observed 
emission features with simple spectral synthesis models 
of LTE emission from gaseous slabs of a given temperature, 
column density, and emitting area (e.g., Carr \& Najita 2008, 2011; 
Salyk et al.\ 2011a). 
Although protoplanetary disks are not isothermal (vertically or
radially) and may not be in vibrational LTE, isothermal slab models
can provide good fits to the data. They also provide estimates of the
physical conditions in the line emitting regions.

For each of the two observed targets,  
we first calculated synthetic spectra to match the IRS SH spectra
and then generated high resolution spectra using the same model 
parameters for comparison with the TEXES data. 
The reduction of the IRS SH data for \as205n\ (2004 August 28, AOR
5646080) and DR Tau (2008 October 8, AOR 27067136) followed the
methods given in Carr \& Najita (2011).  The procedure and molecular
linelists used in the modeling are also described in Carr \& Najita
(2011).
While the molecular emission lines are not resolved with 
IRS ($R$=600), the IRS spectra have the advantage of including 
many lines with different
A-values and energy levels, which constrains the temperature
and column density (e.g., Carr \& Najita 2008, 2011; Salyk et al.\ 2011a).
Thus, the IRS spectra complement the TEXES data, which  
include only a few lines spanning a limited range of excitation. 

In the slab models, the dust continuum is assumed to be negligible in 
the layer of the atmosphere that produces the molecular emission, 
consistent with our expectations for disk atmospheres. 
Recent disk atmosphere models by Najita \& \'Ad\'amkovics (2017), 
which include irradiation by UV continuum and \Lya,
find that warm HCN and water are abundant in the disk atmosphere 
over a vertical column density of $N_H \sim 10^{22}\psqcm$ in 
hydrogen nuclei. 
If we assume that the dust in the atmosphere is reduced by a factor 
of $\sim 100$ compared to ISM conditions, 
as is inferred for T Tauri disks and attributed to grain settling 
(Furlan et al.\ 2006),
the vertical optical depth of the HCN-emitting region is 
$A_V \sim 0.05$ and much lower at mid-infrared wavelengths.

\subsubsection{\as205n}

We first fit the IRS water emission spectrum from \as205n\ using synthetic 
spectra generated with the LTE slab emission model. 
As described above, the IRS spectra have the advantage of 
including many water lines with different
A-values and energy levels, which constrains the temperature
and column density (e.g., Carr \& Najita 2008, 2011).
The temperature and column density were determined by $\chi^2$ fits to
water features between 12\micron\ and 16\micron.
The best fit to the \as205n\ IRS water spectrum (Fig.~18) gives 
a temperature of $680 \pm 80$\,K, 
a column density of $N_\water=1.3$(+0.9/-0.5)\,$\times 10^{18}\psqcm$, 
and an emitting area of $\pi R_e^2$ 
where $R_e=1.90\pm 0.14$\,AU, for a distance of 130 pc
(Mamajek 2008; Loinard et al.\ 2008; Wilking et al.\ 2008).
The parameters 
are given in Table 3.
Note that in fitting the IRS spectra of water, temperature and column
density are correlated (Salyk et al.\ 2011a) such that a higher bound on
temperature corresponds to a lower bound on column density and a smaller
emitting area.

To model the HCN and \ctwohtwo\ Q branch emission 
(at 14\micron\ and 13.7\micron\ respectively),  
we subtracted the synthetic water emission from the IRS spectrum.
While LTE slab models provide a general match to the IRS water spectra,
the fits are far from perfect (Carr \& Najita 2011; Salyk et al.\ 2011a).
Mismatches between the observed and model spectra are likely due to 
a range in temperature and column density in actual disk atmospheres,
non-LTE level populations, and possibly missing transitions in the water
line list or other unidentified features.
Of relevance to the HCN and \ctwohtwo\ modeling, the water model appears to
overcorrect for the few \water\ features that fall within the Q branches.
The location of these \water\ features are marked in Figures 8 and 11.
The pixels affected by these lines were ignored in the modeling.

In modeling the HCN and \ctwohtwo\ emission, we assumed that 
they have the same
emitting area and temperature as the \water\ emission and adjusted
the column density to match the flux in the Q branches.
The assumption of similar emitting areas  
for \water, HCN and \ctwohtwo\
is reasonable, given the similarity of their line profiles in the
TEXES data. 
We find that the HCN and \ctwohtwo\ emission can be matched with  
column densities 
$N_{\rm HCN}=4.2 (\pm 0.2)\times 10^{15}\psqcm$
and $N_{\rm \ctwohtwo}=1.5 (\pm 0.2) \times 10^{15}\psqcm$.
The shape of the Q branches is diagnostic of
the ro-vibrational temperature of the gas. 
Comparison of the observed and model HCN spectra (Fig. 8) shows that HCN is
consistent with having the same $\sim 680$\,K temperature as \water.
To explore the range of possible temperatures for the HCN emission, we held the
emitting radius constant but allowed both temperature and column density to
vary. This gave a best fit at T = 670\,K with a range of 590 to 770\,K,
similar to the temperature range for \water.
For \ctwohtwo, the smaller width of the band and the 
water line at 13.69\micron\  
prevent a definite determination of its temperature. 

The parameters derived from the IRS spectra (Table 3) were then used
to calculate LTE model spectra for comparison to the TEXES data. 
The IRS parameters work well for the TEXES \water\ spectrum:
the line ratios of the 3 best-measured water lines are reproduced
and the line fluxes are within 15\% of the observed values.
For the model plotted in Figure 9, the \water\ column density was increased to
$N_\water=1.6 (\pm 0.1) \times 10^{18}\psqcm$
to match the line fluxes, although a small change
in radius, temperature, or flux calibration would work equally well.
The optical depths for the TEXES \water\ lines are of order unity;
therefore, the ratios of the three lines depend on both
water temperature and column density.
The match of the IRS based model to the TEXES data (Fig.\ 9)
shows that the TEXES and IRS water emission are consistent
with the same LTE temperature, column density, and emitting area.

When the parameters used to fit the IRS spectra of HCN and \ctwohtwo\ 
are applied to the TEXES spectrum (Figure 10; dotted red line),
we find that the HCN lines are under-predicted by 36\%,
while the \ctwohtwo\ line is over-predicted by 44\%.
A change in the joint emitting area of the molecules
could not produce this effect.
The line strengths in the TEXES spectrum can be fit
(solid red line in Fig.\ 10)
by increasing the column density of HCN to
$6.8 (\pm 0.5) \times 10^{15}\psqcm$
and decreasing the column density of \ctwohtwo\ to
$1.0 (\pm 0.1) \times 10^{15}\psqcm$.

\begin{deluxetable}{llllllll}
\tabletypesize{\footnotesize}
\tablecaption{\label{t:std}Parameters For Slab Models}
\tablehead{
Source	 & Spectrum & T  & R     & N(\water) & N(HCN)	& N(\ctwohtwo)	& N(HCN)/N(\water) \\
	 &          & (K)& (AU)  &(cm$^{-2}$)&(cm$^{-2}$)&(cm$^{-2}$) &  }
\startdata 
\as205n\ & IRS   & 680   & 1.90  & 1.3(18) & 4.2(15) & 1.5(15) & 0.0033 \\
\as205n\ & TEXES & 680   & 1.90  & 1.6(18) & 6.8(15) & 1.0(15) & 0.0043 \\
DR Tau 	 & IRS   & 690   & 1.33  & 9.3(17) & 5.7(15) & 6.8(14) & 0.0062 \\
DR Tau 	 & TEXES & 690   & 1.33  & 4.6(17) & 4.4(15) &  ---    & 0.0097 \\
\enddata
\tablecomments{The same temperature and emitting area are assumed 
for all species. For uncertainties, see text.}
\end{deluxetable}

\subsubsection{DR Tau}

The same procedure was followed for DR Tau.
The fit to the IRS water spectrum of DR Tau gave 
T = 690$\pm 75$\,K,
$N_\water=9.3$(+6.7/-3.4)\,$\times 10^{17}\psqcm$, and 
$R_e=1.33\pm 0.07$\,AU,
assuming a distance of 140 pc (Fig.~19 in Appendix B).
After subtracting the model \water\ spectrum, the HCN and \ctwohtwo\ 
Q branch fluxes were fit by adjusting the column densities, assuming
the same emitting area and temperature as for the \water\ emission.
Figure 11 
shows the resulting fit,  
with column densities of  
$N_{\rm HCN}=5.7 (\pm 0.3) \times 10^{15}\psqcm$
and $N_{\rm \ctwohtwo\ }=6.8 (\pm 1.0)\times 10^{14}\psqcm$.
While the \water\ temperature of 690 K is consistent with the shape of the
HCN Q branch, a higher temperature ($\sim 800$\,K)
improves the fit slightly, with an acceptable range of 700 to 1000\,K.
The \ctwohtwo\ Q branch is relatively weak in DR Tau, 
and no conclusion about the \ctwohtwo\ temperature is possible. 

When the water emission parameters derived from the IRS spectra are used 
to model the TEXES spectrum, the water line fluxes are overpredicted
by nearly a factor of two (Fig.~12). 
A reduction of the column density to $N_\water=4.6 (\pm 0.2)\times 10^{17}\psqcm$
matches the line fluxes.
The temperature could also be lowered to reduce the line fluxes,
but then the observed line ratios no longer match.
The model \water\ fluxes can also be reduced by decreasing the emitting area,
or increasing the continuum flux (ie., the flux calibration), but this would
cause the HCN flux to be under-predicted.
With the nominal IRS-derived model, the HCN line is over predicted by
$\sim 30$\%.
To reproduce the observed HCN line strength (Fig. 13) requires decreasing
the HCN column density to $4.4 (\pm 0.8)\times 10^{15}\psqcm$.
The prediction for the \ctwohtwo\ R21 line (not shown) is 
consistent with the non-detecton of this line in the TEXES data.

\subsubsection{Variability}

As described above, we find changes in the derived water and HCN column
densities between the IRS and TEXES observations for both \as205n\ and
DR Tau (see Table 3). 
Because the TEXES observations were flux calibrated by adopting the
continuum flux from the IRS spectra,
any variation in the continum level would produce apparent changes
in line flux and the derived column density.
However, a simple flux calibration scaling alone cannot explain all the
observed changes, because some of the discrepancies between model and
observed line fluxes differ in magnitude and sometimes in
direction.
Nevertheless, it is possible to compare the ratios of column densities.
If the continuum spectral shape of the source is constant
over the small range of TEXES wavelengths studied,
the equivalent widths of the HCN and water lines will accurately
reflect their relative fluxes.

Based on the above LTE slab modeling of the IRS spectra,
the HCN/\water\ column density ratio is 
0.0033 ($+$0.0021/$-$0.0013) for \as205n\  
and 0.0062 ($+$0.0033/ $-$0.0023) for DR Tau.
These uncertainties come from the range in the ratio
obtained at the extremes of the confidence boundary
in temperature and column density parameter space for the 
fit to the IRS water spectrum
and are dominated by the allowed range in the water 
column density. 

To compare the HCN/\water\ column density ratios from the IRS and TEXES data, 
we calculate the uncertainty in the IRS ratio using the error in the 
IRS water column at a fixed temperature.
This choice is consistent with our use of the temperature and $R_e$ 
from the IRS water spectrum in fitting the other data (Table 3) and 
corresponds to a $\sim 20$\% error in $N(\water)$ for both \as205n\  and DR Tau.
The other uncertainties that matter are the measurement uncertainties
in the IRS HCN band flux and the TEXES line fluxes.

Including these uncertainties in comparing the HCN/\water\ ratios,
we obtain for \as205n\  a column density ratio 
HCN/\water\ = $0.0033 \pm 0.0007$ from the IRS data 
and $0.0043 \pm 0.0004$ from the TEXES data.
For DR Tau, the ratios are $0.0062 \pm 0.0013$ from the IRS data 
and $0.0097 \pm 0.0018$ from the TEXES data.
While the change in the HCN/\water\ ratio is 30\% for \as205n\  and 56\% for DR Tau,
the significance in the difference is modest, 
$1.3 \sigma$ and $1.6 \sigma$, respectively,
providing 
marginal evidence for variation in the column density ratio.

One of the limitations in this comparison is the presumption that the
temperature and emitting area are the same for both molecules at both epochs
of observation.
We found that the IRS spectra are consistent with the same temperature
for \water\ and HCN, within the uncertainties. Furthermore, the water line ratios
in the TEXES spectra are consistent with the IRS derived temperature for \water.
However, there is no independent constraint on the HCN temperature for the
TEXES observation. The presumption of the same emitting area is based on the
similarities of the \water\ and HCN velocity profiles, but we can only assume that
the emitting areas were the same at the time of the IRS observations.

Another limitation
is that we are comparing models for low-resolution (IRS) observations of
the HCN Q branch, which is a blend of ro-vibrational lines dominated by the
v$_2$=1--0 and 2--1 bands,
with observations of individual lines (TEXES) in the v$_2$=1--0 R-branch.
The upper energy level of the R-branch lines observed with TEXES is close to the
flux weighted average of lines within the observed Q branch.
However, if non-LTE effects were to produce different rotational and vibrational
excitation temperatures, then a temperature that fits the entire Q branch may
not be the appropriate excitation temperature for the observed v$_2$=1--0 lines.

All of the above assumptions can eventually be tested with multiple epochs of
high-resolution spectroscopy of water and HCN that cover more transitions.
By measuring a broader range in HCN line excitation,
and with sufficient S/N to measure v$_2$=2--1 transitions,
the HCN rotational and vibrational temperatures can be determined.

\section{Discussion}

\subsection{\it Emission from the Inner Disk} 

We find that the line profiles of the MIR HCN and \water\ emission 
from DR Tau and \as205n\ are roughly symmetric and centered 
within a few $\kms$ of the stellar velocity (Appendix A). 
The emission also overlaps in velocity 
other known molecular emission diagnostics 
from the inner disk regions of these sources
(rovibrational CO, \water, UV fluorescent \hm; 
e.g., Salyk et al.\ 2008, 2011b; 
Schindhelm et al.\ 2012; 
Banzatti et al.\ 2014; 
L.\ Brown et al.\ 2013; 
J.\ Brown et al. 2013). 
These properties are consistent with emission from a rotating
disk and not with high velocity outflows or inflows. 
Magnetocentrifugal winds and magnetospheric infall 
produce strongly blue- or red-shifted emission 
and/or absorption features extending to $\sim 100\kms$ or more 
even at the relatively low inclination of our sources, 
e.g., forbidden optical line emission from winds 
(Hartigan et al.\ 1995; Simon et al.\ 2016) and 
magnetospheric infall emission/absorption 
profiles observed in HI 
(e.g., Edwards et al.\ 1994; Muzerolle et al.\ 1998a,b; 
Folha \& Emerson 2001). 

Millimeter imaging and infrared spectroastrometric studies 
point to a low inclination for \as205n\ ($i=15^\circ$; as discussed in 
Salyk et al.\ 2014). 
Inclination estimates range from low to high for DR Tau, 
with spectroastrometric studies of the inner disk 
favoring a low inclination. 
The Pontoppidan et al.\ (2011) study of CO rovibrational emission 
found a good fit to their data assuming an inclination of 9 degrees.  
L.\ Brown et al.\ (2013) were able to fit their spectroastrometric 
$L$-band water observations with an inclination of $i=13^\circ$. 

Photoevaporative disk winds are expected to produce observable 
blueshifts of 5--10\kms\ at these low inclinations 
(Font et al.\ 2004; Alexander 2008). 
In contrast, the water profiles  
of both sources studied here show, if anything, small redshifts 
rather than blueshifts. 
The water profile of DR Tau is $+3\kms$ from the average 
of the reported stellar velocities 
(Petrov et al.\ 2011; Nguyen et al.\ 2012), and  
the \as205n\ profile is shifted by $+4\kms$ 
(Melo 2003; Appendix A). 
The absence of a photoevaporative signature in the 
HCN and \water\ profiles is not surprising. 
Driven from the disk surface 
by X-ray and UV irradiation, photoevaporative flows are 
likely to be atomic or ionized rather than molecular and unlikely 
to contribute to the molecular emission profile.
Thus, the line profiles of the HCN and \water\ emission are best 
explained as arising in the inner disk. 

The resolved line profiles also constrain the range of disk radii 
from which the emission arises. 
AU-sized projected emitting areas have been inferred 
for the MIR molecular emission from CTTS based on 
{\it Spitzer} spectra, which are spatially and spectrally unresolved 
(Carr \& Najita 2008, 2011; Salyk et al.\ 2011a).
From simple slab emission models, we infer 
water emitting areas $\pi R_e^2$ with $R_e$ of 
1.3\,AU for DR Tau and 1.9\,AU for \as205n\ (\S3.2),
similar to values previously reported in the literature. 
Salyk et al.\ (2011a) previously inferred water emitting areas 
$\pi R_e^2$ with $R_e$ of 1.1\,AU for DR Tau and 
2.1\,AU for \as205n\ using a similar approach.

Using the resolved line profiles, we can obtain an 
independent constraint on the range of disk radii 
responsible for the emission given the disk inclination 
and stellar mass for each source. 
For \as205n, with a stellar mass of $\Mstar = 1.1\Msun$ 
(Najita et al.\ 2015) and an inclination of $i=15^\circ,$ 
the observed $\sim 30\kms$ HWZI of the \water\ emission 
corresponds to an inner emission radius of 0.07\,AU. 
Emission from gas at 1.7\,AU would have a 
projected velocity of $v\sin i = 6.2\kms$, 
similar to the half-width of the flat-topped portion 
of the water line profile (Fig.~6). 
Thus, the line profile is consistent with \water\ emission 
extending from 0.07\,AU to $\sim 2$\,AU. 
The HCN profile of \as205n\ is similar to the \water\ profile 
at low velocities, indicating that the HCN emission 
extends over a similar range of radii.
Because of their lower signal-to-noise ratio, it is not possible 
to determine whether the wings of the HCN profile have the 
same velocity extent as the \water\ profile. 
The HCN emission extends to at least $\pm 15\kms$
from line center, i.e., inward in radius to at least 0.3\,AU.

For DR Tau, with a stellar mass of $\Mstar = 1.2\Msun$ 
(Andrews et al.\ 2013 using stellar evolutionary tracks 
from Siess et al.\  2000) 
and $i=8^\circ$, the observed HWZI of at least $20\kms$ for the \water\ lines 
corresponds to an inner emission radius of 0.05\,AU or smaller. 
Emission from gas at 1.4\,AU would have a projected velocity 
of $v\sin i = 4\kms$. 
We observe \water\ emission throughout 
this range of velocities, and the strongly peaked profile suggests that 
the emission may extend beyond 1.4\,AU.
Here we have adopted a lower inclination ($i=8^\circ$) than the values 
preferred by Pontoppidan et al.\ (2011; $i=9^\circ$, $\Mstar=1.0\Msun$) 
and L.\ Brown et al.\ (2013; $i=13^\circ$; $\Mstar=0.4\Msun$) 
to compensate for the larger stellar mass adopted here.

\subsection{\it HCN and Water Probe the Same Disk Volume}

The similar line profiles and emission temperatures for 
HCN and water suggest that the two diagnostics probe the 
same volume of the disk atmosphere. 
Firstly, the HCN and \water\ profiles extend 
over approximately the same range of velocities 
in the case of both \as205n\ and DR Tau, 
consistent with the HCN and water emission arising 
from approximately the same range of disk radii in 
both systems.
Secondly, the similar emission temperatures for HCN and water 
are consistent with the HCN and water emission arising from the 
same vertical height in the disk. 
Models of inner disk atmospheres 
irradiated by stellar UV and X-rays 
find that the gas temperature exceeds the dust 
temperature in the atmosphere and transitions from a 
hot ($\sim 4000$\,K) atomic region at the top of the atmosphere 
to a warm molecular region ($\sim 300-1000$\,K) at 
intermediate heights, before reaching the dust temperature 
deeper in the atmosphere 
(Adamkovics et al.\ 2014; Najita et al.\ 2011; 
Najita \& Adamkovics 2017). 
Both HCN and water are abundant in the 
intermediate layer, the warm molecular region, 
consistent with the similar line profiles and temperatures 
found here for the HCN and water emission.

The inferred temperatures and column densities of 
the HCN and water emission are also roughly consistent 
with thermal-chemical models of irradiated disk atmospheres. 
In comparison with the $\sim 700$\,K temperature and 
$R_e \sim 1.5$\,AU emitting area found for the HCN 
and \water\ emission, 
the reference model of Najita \& Adamkovics (2017) 
for a generic T Tauri disk atmosphere 
irradiated by stellar FUV continuum, \Lya, and X-rays, 
is characterized at 1\,AU by vertically coincident 
warm (300--1000\,K) HCN and water with 
vertical column densities 
($3\times 10^{15}\psqcm$ and $2\times 10^{17}\psqcm,$ respectively) 
similar to the measured line-of-sight columns 
($\sim 5\times 10^{15}\psqcm$ and $\sim 1\times 10^{18}\psqcm,$ respectively).  
Closer agreement may be possible with disk atmospheres 
more closely matched to the star+disk properties of the sources 
studied here.

\subsection{\it Chemical Fingerprint of Planetesimal Formation}

Our results support the interpretation of the observed 
trend of increasing HCN/\water\ emission with disk mass 
as a possible chemical fingerprint of planetesimal formation 
and core accretion in action 
(Najita et al.\ 2013; Carr \& Najita 2011). 
The core accretion model of planet formation is compelling 
in its ability to account for planet formation outcomes 
such as the existence of 
small rocky planets as well as ice and gas giants  
within the Solar System and in the exoplanet population. 
But what does core accretion {\it look} like? 
Can any observations reveal whether known 
protoplanetary disks are actually engaged in core accretion? 
One approach is to look for observational evidence of the 
formation of planetesimals and protoplanets, which play 
no role in a competing theory of planet formation such as 
gravitational instability.  

Observationally elusive, planetesimals are difficult to 
detect directly because of their small size, small mass, 
and intrinsic faintness. 
However, they may potentially reveal themselves through 
the chemical signature they induce in the gaseous disk 
within the snowline. 
When icy objects grow large enough (kilometer-sized or larger) 
to decouple from inward migration induced by gas drag, 
they are sequestered in the icy region of the disk beyond the 
snow line (beyond $\sim 1$\,AU; Ciesla \& Cuzzi 2006). 
The gas that accretes into the warm inner disk region within the 
snow line is thereby depleted of water (and oxygen) resulting in 
a higher C/O ratio than the disk overall. 
Thermal-chemical models of disk atmospheres predict that 
only a modest enhancement in 
the C/O ratio of the inner disk (factor of 2) is sufficient to 
induce a large, observable change in the HCN/\water\ ratio in the 
atmosphere (factor of 10), 
consistent with the range of observed molecular emission 
flux ratios (Najita et al.\ 2011). 

As a result, molecular abundances in the inner disk (within the snow line) 
may encode evidence for icy planetesimal formation 
beyond the snowline. 
Because disks with higher masses are expected to grow planetesimals 
more rapidly, we expect to see the HCN/\water\ ratio in the inner disk 
increase with disk mass in systems undergoing planetesimal formation, 
a trend that is indeed observed (Najita et al.\ 2013; see also Fig.\ 14). 
If this interpretation of the observed trend is correct, it implies that 
{\it most} T Tauri disks are in an advanced stage of planet formation, 
having built and sequestered large icy bodies beyond the snow line 
(Najita et al.\ 2013).
This 
interpretation assumes that the HCN and \water\ emission 
arise from the same volume of the disk atmosphere. 
In finding that mid-infrared HCN and water emission probe the 
same volume of the disk atmosphere, our results support the 
hypothesis that the trend 
between the HCN/\water\ flux ratio and disk mass is evidence 
for planetesimal formation. 

The inference that planet formation is well advanced in most 
T Tauri disks is supported by a comparison of 
the mass of small solids present in T Tauri disks 
(the dust disk mass distribution) with 
the solids locked up in known exoplanet populations 
(Najita \& Kenyon 2014). 
The comparison  shows that there are fewer small solids in T Tauri disks 
than in the known exoplanets.  
If T Tauri disks are the birthplaces of planets, 
their low inventory of solids implies that 
they are by no means unevolved ``primordial disks.'' 
More likely, most of the solids in most 
T Tauri disks have already grown into large sizes that are 
invisible to millimeter observations (beyond cm sizes) 
and/or they have been  
concentrated at small radii (within a few AU) where they are 
optically thick at millimeter wavelengths. 
The trend of HCN/\water\ vs.\ disk mass goes a step further and 
implies that the solids beyond the snow line 
have grown much beyond cm size and large enough to decouple 
from gas drag (i.e., kilometer-sized or larger). 

This picture is consistent with the short timescale inferred 
for planetesimal formation in the Solar System. 
Radiometric analyses of meteorites 
find that differentiated planetesimals accumulated soon 
after the formation of CAIs (calcium aluminum inclusions), 
i.e., that planetesimal formation was well underway at the 
onset of the T Tauri epoch of Solar System history 
(Kleine et al.\ 2009; Dauphas \& Chaussidon 2011).  

Figure 14 plots as a function of disk mass 
the HCN/\water\ flux ratio 
measured from IRS spectra for the sources reported 
in Najita et al.\ (2013; red diamonds) 
and the sources studied here (blue diamonds). 
The HCN flux shown includes a correction for water emission 
blended with the HCN Q branch. The water flux shown is the 
sum of the water emission 
features at 17.12\micron, 17.22\micron, and 17.36\micron, 
and disk masses are derived from submillimeter continuum 
emission (Najita et al.\ 2013, 2015). 

The trend of HCN/\water\ vs.\ disk mass, while clearly apparent, 
has significant scatter.
Multiple factors may be responsible for the scatter.
The extent of planetesimal formation in the giant planet region 
may differ between sources of similar outer disk mass. 
Excitation effects may reduce HCN emission in inner disks with 
lower densities (\S 4.5).  
HCN and \water\ may not always probe the same volume. 

Any time variability in the HCN/\water\ emission ratio 
(\S 3.2.3) will also contribute to the observed scatter. 
Figure 14 shows 
the larger HCN/\water\ ratios we would infer from 
the TEXES data (blue circles) if the 
larger HCN/\water\ column density ratios found for 
the TEXES data (compared to the IRS data) reflect 
time variability (\S 3.2.3) and if the change in 
the inferred column density ratio (Table 3)
translates into a change in the HCN/\water\ flux ratio. 
Accordingly, the flux ratios of \as205n\ and DR Tau in the TEXES data 
are shown enhanced by factors of 1.3 and 1.6, respectively, 
over their IRS flux values. 
The magnitude of the variability in the HCN/\water\ flux ratio 
found for \as205n\ and DR Tau is similar to the magnitude of the scatter
observed in the relationship of HCN/\water\ versus disk mass.
(The overall trend of HCN/\water\ flux ratio 
with disk mass extends, of course, over a much larger 
range of values.) 
Further study of variability in the HCN/\water\ ratio 
is needed to understand whether this range of variability is 
typical of T Tauri disks. 

Changes in molecular emission from the inner disk could 
arise from changes in the instantaneous UV field 
(driven by variations in stellar accretion), 
instantaneous local accretion heating
(e.g., from time-varying energy dissipation due to the magnetorotational
instability), grain settling or
dust-lofting events in the disk atmosphere (e.g., due to disk
turbulence), or other effects. 
These processes affect the photochemistry of the disk
atmosphere; the depth of the warm disk atmosphere that can produce
MIR molecular emission; and the depth to which we can see into the
atmosphere. Because the HCN emission tends to be concentrated toward
the top of atmosphere, whereas water emission can extend deeper
(e.g., Najita \& Adamkovics 2017), these processes can enhance
or reduce the HCN/\water\ flux ratio.

A concrete illustration of the role of one process, accretion
heating, in altering the molecular emission ratio comes from
observations of the extreme outburst observed from EX Lupi (Banzatti
et al.\ 2012). The {\it Spitzer} water emission increased
dramatically in outburst, while emission from the organics completely
disappeared. More modest accretion variability may produce more
modest changes in the relative molecular fluxes.  

\subsection{\it Inner Disk Molecular Abundances}

Our constraints on the range of disk emission radii 
for HCN and water also bear on the abundance ratios 
inferred for inner disks. Molecular abundances 
can probe the extent to which inner disks are 
chemically active, i.e., whether inner disks actively 
synthesize molecules or primarily inherit their 
abundances from the ISM or larger disk radii. 
As noted in \S 1, some earlier analyses of spectrally unresolved data 
inferred that inner disks have an HCN/water abundance that is  
significantly enhanced, by about an order of magnitude, relative 
to that of comets (Carr \& Najita 2011, 2008).
If comets are a surrogate probe of the chemical conditions in the 
giant planet region of the disk, where they are believed to have 
formed, the difference between the abundances of inner disks and 
comets can be interpreted as evidence for an active 
inner disk chemistry. 

The low spectral resolution of {\it Spitzer}/IRS introduces potential
uncertainty in the inferred column density ratios and abundances, 
because lines are blended and the spatial information 
from resolved line profiles is unavailable. 
In general, there can be signficant degeneracies between model 
parameters, even for LTE slab models 
(Salyk et al.\ 2011a; Carr \& Najita 2011). 
Modeling assumptions may also differ.

In their modeling of {\it Spitzer} IRS spectra,  
Carr \& Najita (2008, 2011) favored a fit with 
a smaller emitting area for HCN than water. 
In their minimum chi-square fits, 
in which the HCN emission is close to optically thick, 
the HCN emitting area is a factor of 3--12 smaller than the 
water emitting area, and the HCN/\water\ column density 
ratio is high (0.03--0.1; Fig.\ 15, open red squares). 
In contrast, under the assumption of equal emitting areas for 
HCN and water, 
the HCN emission is optically thin, and 
the HCN/\water\ column density ratios are 4--30 times smaller 
than in the best fit, but still fall within the 
confidence interval of fits to the HCN bandhead 
(see Carr \& Najita 2011; Fig.\ 15, solid red squares). 

Salyk et al.\ (2011a) also reported molecular
column densities from simple slab fits to {\it Spitzer} 
IRS spectra under the assumption of equal emitting areas 
for all molecules. 
In contrast to the smaller wavelength region studied by 
Carr \& Najita (2011; 12--16\micron), 
Salyk et al.\ (2011a) fit the 
strengths of water emission features from a much 
broader range of wavelengths (10 to 35\micron), among 
other differences in the modeling procedure. 
As shown in Figure 15, the molecular column density ratios 
reported by Salyk et al.\ (magenta squares) do not overlap 
those of Carr \& Najita (2011; solid red squares), illustrating how 
different modeling choices 
(e.g., wavelength region studied, assumed line broadening, 
the weight given to individual spectral features)
can lead to different results, even in the context 
of simple slab models. 
In general, Salyk et al.\ found lower emission 
temperatures and higher column densities for water, 
a result that pulls the molecular column density ratios 
in Figure 15 toward the lower left of the plot. 

Now with velocity-resolved spectra of water and HCN 
in hand---data that directly constrain molecular 
emitting radii and resolve line blends---we have, 
for the first time, an opportunity 
to directly assess some of the degeneracies in modeling
{\it Spitzer} IRS spectra.
For the two sources studied here,
we find that the high-resolution water spectra 
are consistent with the
high-temperature, low-column-density interpretation of the 
{\it Spitzer} water emission 
(see also Carr et al.\ 2018, in preparation).
Similarly, we find that \as205n\ and DR Tau have approximately 
equal HCN and water emitting areas, which helps to distinguish 
between high- and low-column-density solutions for the HCN 
emission as well. 
Assuming the {\it Spitzer} HCN emission temperature applies 
to the TEXES HCN emission, 
the {\it Spitzer} and TEXES spectra can be fit with similar
HCN/\water\ column density ratios.

The Spitzer HCN/\water\ 
column density
ratios (Fig.\ 15, red diamonds) are 0.0033 for \as205n\ and 0.0062
for DR Tau; the TEXES HCN/\water\ column density ratios 
(Fig.\ 15, large red circles with black dots) 
are 0.0043 for \as205n\ and 0.0097
for DR Tau (3.2; Table 3).  These values agree with the low end 
of the HCN/\water\ column density ratios reported by 
Carr \& Najita (2011) and are closer to
the average value of comets (blue symbols).
For comparison, the range of molecular abundances measured 
for hot cores is also shown in Figure 15 (green cross).

These results endorse some aspects of earlier work and also 
suggest the need to extend our study to a wider variety of 
sources. 
We confirm for the two sources studied here
that simple slab fits to their IRS spectra 
(following the methodology of Carr \& Najita 2011) are 
consistent with the properties inferred from 
their TEXES spectra (\S 3.2). These results 
suggest a measure of confidence in our ability 
to infer abundances from lower resolution data. 
If the equal HCN and water emitting areas we find 
for the sources studied here 
also apply to the sources studied by Carr \& Najita (2011), 
their HCN/\water\ column density ratios would be lower than 
the favored best-fit values and, thereby, alter our 
view of the level of chemical activity in inner disks. 
However, the results obtained here may not apply
to other classical T Tauri stars, because \as205n\ and DR Tau 
are both very active sources with high accretion rates. 
High spectral resolution observations, similar to those 
presented here, but of more typical T Tauri 
stars (like those studied in Carr \& Najita 2011),  
are likely critical in order to measure with confidence molecular 
column density ratios that can be usefully compared to 
the abundances of comets and hot cores.

\subsection{\it Excitation of HCN}

Finally, our results bear on models for the excitation of HCN. 
Bruderer et al.\ (2015) studied the possibility of non-LTE excitation 
of HCN, including excitation by collisions with \hm\ and 
radiative pumping by infrared photons. 
They concluded that infrared pumping is
the major excitation mechanism for ro-vibrational HCN emission. 
This was particularly true for the 3\micron\ lines of the $\nu_1$ band,
because the high critical densities of these transitions 
($n_{\hm} \sim 10^{13}-10^{14}\pccm$) are only present in their disk model
well below the dust photosphere of the inner disk;
the excitation of these transitions above the disk photosphere 
therefore relies on radiative pumping.
In contrast, the critical density of the 14\micron\ lines of the 
$\nu_2$ band is lower ($\sim 10^{11}\pccm$), 
and collisional excitation could affect the 14\micron\ emission
from the inner 2\,AU of their disk model. 

Using a disk model calculated specifically for the case of \as205n,
Bruderer et al.\ predicted 
3\micron\ and 14\micron\ line profiles and intensities
for the HCN emission from the disk.
They found that the 3\micron\ and 14\micron\ lines can have 
substantially different line profiles and 
that infrared pumping can excite the transitions 
in both bands out to large radii ($\sim 10$\,AU), producing 
very narrow line profiles.

To test these predictions, 
we compared the HCN line profiles of \as205n\ measured with 
TEXES in the mid-infrared 
with the 3\micron\ line profiles from CRIRES 
(Mandell et al.\ 2012).
We obtained an average profile for the 3\micron\ lines by 
averaging together the P5, P11, P12, and P13 lines, the cleanest
HCN lines in the spectrum.
The average 3\micron\ profile is compared to our average profile for the
MIR lines in Figure 16. 
The MIR and 3\micron\ HCN profiles 
are remarkably similar.

The velocity width for both line profiles (FWHM $\sim 20\kms$) is
much larger than the $\sim 10\kms$ predicted by the Bruderer et al.\
models in which HCN is excited by infrared pumping out to $\sim 10$\,AU.
The discrepancy implies 
that either HCN is not abundant beyond a couple of AU
or infrared pumping does not dominate at these radii.
The Bruderer model that best fits our observed width for the 
mid-infrared lines is the ``jump'' model (see their Fig.\ 10) 
in which the HCN 
abundance is assumed to be greatly reduced beyond $\sim 2$\,AU. 
The radius of the abundance decrement is consistent with the
emitting area derived from our modeling of the IRS spectra (\S 3.2). 
However, the same jump model also predicts a 3\micron\ line that is
enhanced by radiative pumping and is much broader 
($\sim 45\kms$) than observed.

Collisional excitation is most likely the dominant excitation mechanism
for the 14\micron\ HCN band. 
From fitting the Q branch at 14\micron\ (\S 3.2.1), 
we know that higher vibrational states, up to at least v$_2=3$,
must be populated. Populating these vibrational states via infrared
pumping would be difficult because of their weak radiative connection
to the ground state.
In the Bruderer et al.\ models, collisional excitation was found to be important
for the 14\micron\ lines in the inner 2 AU of the disk, consistent with our
observed line profiles.
The critical density of the HCN lines we observed with TEXES is 
$\sim 10^{11}\pccm$, 
not much higher than those of the observed rotational water lines 
($3\times 10^{10} - 6\times 10^{10}\pccm$)
and similar to the density in disk atmosphere models where HCN is abundant
(Adamkovics et al.\ 2014; Najita \& Adamkovics 2017).
All of this argues that the 14\micron\ transitions could be close to LTE.

The role of radiative pumping in exciting the 3\micron\ and MIR 
HCN lines 
is unclear, as the specific line profiles predicted by Bruderer et al.\
are not confirmed by the observations. 
The 3\micron\ profile is indistinguishable from the MIR profile, 
showing that they form over the same range of radii and
suggesting that the two bands could share the same excitation mechanism.
If radiative pumping does play a role, the details of how the 
radiative excitation occurs 
must differ from the description put forth in Bruderer et al.\ (2015).

In addition to comparing the line profiles,
we can also compare the expected LTE emission line fluxes with 
the observed fluxes.
Using the LTE slab model that we used to fit the Q branch of
the 14\micron\ band in the IRS spectrum, we  
predicted the strength of the 3\micron\ HCN lines 
and compared it with the 3\micron\ line strength from 
Mandell et al.\ (2012; Figure 17).
The simple LTE model produces a good match to the 3\micron\
HCN line flux in the CRIRES spectrum,
better than the same model applied to the TEXES data (Fig.\ 12). 
One should note that neither the CRIRES spectrum nor the TEXES spectrum
were independently flux calibrated, and both these and the 
{\it Spitzer} IRS spectrum were observed years apart.
Nevertheless, the intensity of the 3\micron\ and MIR lines, 
as studied here, are consistent with HCN being close to 
LTE.

The similar 3\micron\ and MIR line profiles for \as205n, the
warm vibrational temperature ($\gtrsim 700$\,K) implied by the fit
to the 14\micron\ band in the IRS spectra, and the flux ratio of
the 3\micron\ and MIR lines, are all consistent with collisional
exciation of the HCN within 1--2 AU of the star.  However, a 
significant uncertainty
is whether the high densities required for collisional excitation
of the 3\micron\ lines 
($\sim 10^{13}-10^{14}\pccm$; see Bruderer et al.\ 2015
Fig.\ 4) are present in the region of the disk atmosphere where the
HCN emission forms or whether a different excitation path is
required to understand the observed 3\micron\ HCN transitions.

One possibility is that HCN is excited by collisions with atomic 
hydrogen rather than \hm. In the disk atmosphere models of 
Najita \& \'Ad\'amkovics (2017), atomic H and \hm\ are comparable 
in abundance in the region where warm HCN is abundant. 
For CO and SiO, the cross sections for collisions with atomic H 
are considerably larger than for those with \hm.  
If the same is true for HCN, it would help to explain the apparent 
LTE excitation of HCN indicated by the results presented here.

\section{Summary} 

The MIR molecular emission from inner protoplanetary disks, detected commonly
in the {\it Spitzer} IRS spectra of classical T Tauri stars, may
provide valuable clues to our understanding of chemical processing
in disks and their planet formation status. Previous studies have
suggested that comparing inner disk abundances with cometary
abundances can probe chemical processing in the terrestrial planet
region of the disk. Other studies have suggested that inner disk
abundances can potentially provide a chemical fingerprint of
planetesimal formation in the giant planet region of the disk, an
observationally elusive process that is a potential signature of
core accretion in action.

The high resolution spectra of MIR water and HCN presented here
supports these ideas. The similar line profiles and emission
temperatures of HCN and water argue that the two diagnostics arise
in the same volume of the inner disk atmosphere. As a result, it
seems plausible that their flux ratios and column density ratios
can constrain the molecular abundance ratio of the inner disk. Thus 
our results support the inference from previous studies that planet
formation is well advanced in most T Tauri disks, with the majority
of disk solids having grown into large, non-migrating objects
(planetesimals or protoplanets) that are sequestered beyond the inner
disk region.

The implications of our results for inner disk abundances is less
clear. The good agreement between the abundance ratios we infer
from the {\it Spitzer} and TEXES data for our two sources suggest
confidence in the abundance ratios inferred from the lower resolution
{\it Spitzer} data. However, if most T Tauri stars have similar
emitting areas of HCN and water, as found here, the true HCN/\water\
abundance ratios are lower than reported by Carr \& Najita (2011).
Because \as205n\ and DR Tau are unusual, very active T Tauri stars,
high resolution spectroscopy of more typical T Tauri stars is needed
to resolve this issue.

\bigskip

\appendix {\bf Appendix A. Comparison of MIR Emission Line Velocities with 
Literature Values}

The line center velocities of the molecular emission from 
DR Tau and \as205n\ are consistent, within the TEXES wavelength 
calibration uncertainty,  
with that of other IR and submillimeter molecular emission from 
the sources and with their stellar velocities. 
Like the MIR water line profiles,  
the 4.5\micron\ CO emission from DR Tau 
is centrally peaked with a FWHM of 15\kms\ and centered at 
$v_{\rm helio} \simeq 25 \kms$ (Bast et al.\ 2011; 
J.\ Brown et al.\ 2013).
The MIR water emission is also centered within $2\kms$ of the 
CO J=3--2 and J=5--6 emission from the outer disk 
($v_{\rm helio}=23\kms$; Thi et al.\ 2001; Greaves et al.\ 2004). 
Petrov et al.\ (2011) found that the stellar velocity of 
DR Tau varies by $\pm 1\kms$ about $v_{\rm helio}=23\pm 2\kms$ 
on a timescale of days,  behavior that they attribute to the 
effect of a rotating emission hot spot on the star. 
Nguyen et al.\ (2012) reported a stellar velocity of $v_{\rm helio}=21.1\kms$
based on a smaller number of measurements (5 epochs).\footnote{Ardila 
et al.\ (2013) cites a stellar velocity of $27.6\kms$ for 
DR Tau, which is attributed to Alencar \& Basri (2000)  
although that value does not actually appear in the latter paper.}  
Thus, the MIR water emission from DR Tau is within 
$3\kms$ of the average of the two reported stellar velocities.

Similar to the situation for DR Tau, the velocity of the 
MIR emission from \as205n\ 
is consistent with that of other molecular emission from \as205n\ and 
with the stellar velocity.   
The $v_{\rm helio}= -5\kms$ velocity centroid of the 
TEXES \water\ emission is consistent 
with the $M$-band CO velocity (Bast et al.\ 2011; J.\ Brown et al.\ 2013)
within the TEXES wavelength calibration uncertainty. 
It is also within $2\kms$ of the millimeter CO emission 
from the outer disk ($v_{\rm helio}=-7\kms$; Salyk et al.\ 2014). 
We find more of a discrepancy with the MIR water emission 
reported by Pontoppidan et al.\ (2010a). 
Their observations, which include 2 of the 3 
higher excitation \water\ lines studied here (806\pcm, 807\pcm),  
show velocity shifts relative 
to CO that vary from line to line, with some lines showing a 
$\sim 5\kms$ blueshift relative to CO.
The lower signal-to-noise ratio of their data may account for 
some of the observed differences. 
The TEXES \water\ emission is close to ($\sim 4\kms$ redward of)  
the stellar velocity. 
The average stellar velocity reported by Melo (2003)
from multi-epoch data is $v_{\rm helio}=-9.4\kms$ 
compared with $v_{\rm helio}=-5.0\kms$ here. 
Other reported stellar velocities for \as205n\ were based on 
observations made at a single epoch. 
Eisner et al.\ (2005) reported $-11.6\kms,$
and L. Prato measured $-7\kms$ (private communication 2015). 

Thus we find that the MIR water emission velocities of \as205n\ and 
DR Tau are within a few \kms\ of 
the stellar velocity and possibly redshifted. 
While the reason for the velocity difference is unclear, 
the situation is not unique.  Bast et al.\ (2011) also reported 
velocity differences of $\pm 5\kms$ 
between stellar velocities and molecular emission (their Fig.~14).
The differences may arise, in part, from  
variations in the stellar radial velocity produced by 
brightness inhomogeneities on the stellar surface or 
orbital motion induced by a close companion.  
As noted above, Petrov et al.\ (2011) attributed the photospheric 
radial velocity variation of DR Tau to the effect of a rotating 
emission hot spot on the star, 
because the radial velocity measured for photospheric lines is 
anticorrelated with the radial velocity variation of chromospheric 
emission lines. 
A close-in giant planet can induce radial velocity variations  
of $\sim 1\kms$ in amplitude (e.g., a stellar motion of this order is 
attributed to the influence of an $8\,M_J$ companion to CI Tau; 
Johns-Krull et al.\ 2016).

\bigskip

\appendix {\bf Appendix B. Supplementary Figures}

Figures 18 and 19 compare the {\it Spitzer} IRS spectra of \as205n\ and 
DR Tau with simple slab models that assume LTE populations. 
The models and parameters are described in greater detail in 
\S 3.2.

\acknowledgements
We thank Avi Mandell for making the $L$-band spectra available. 
This work is based on observations obtained 
through programs GN-2013B-Q-34 and GN-2014B-Q-67 
at the Gemini Observatory, which is
operated by the Association of Universities for Research in Astronomy,
Inc., under a cooperative agreement with the NSF on behalf of the
Gemini partnership: the National Science Foundation (United States),
the National Research Council (Canada), CONICYT (Chile), Ministerio
de Ciencia, Tecnolog\'{i}a e Innovaci\'{o}n Productiva (Argentina),
and Minist\'{e}rio da Ci\^{e}ncia, Tecnologia e Inova\c{c}\~{a}o
(Brazil).  
Basic research in infrared astronomy at the Naval Research
Laboratory is supported by 6.1 base funding. J.\ N.\ acknowledges the
stimulating research environment supported by NASA Agreement No.\
NNX15AD94G to the ``Earths in Other Solar Systems'' program.

\newpage

\noindent {\bf References}

\noindent \'Ad\'amkovics, M., Glassgold, A.\ E., \& 
Najita, J.\ R.\ 2014, ApJ, 786, 135

\noindent Alencar, S. H. P., \& Basri, G. 2000, AJ, 119, 1881

\noindent Alexander, R.\ D.\ 2008, MNRAS, 391, L64. 

\noindent Andrews, S.\ M., Wilner, D.\ J., Hughes, A.\ M., 
Qi, C., \& Dullemond, C.\ P.\ 2009, ApJ, 700, 1502

\noindent Andrews, S.\ M., Rosenfeld, K.\ A., Kraus, A.\ L., 
\& Wilner, D.\ J.\ 2013, ApJ, 771, 129 

\noindent Ardila, D. R., Herczeg, G., Johns-Krull, C. M., et al.\ 2013, ApJS, 207, 1

\noindent Banzatti, A., Meyer, M. R., Bruderer, S. et al.\ 2012, ApJ, 745, 90

\noindent Banzatti, A.\ \& Pontoppidan, K.\ M.\ 2015, ApJ, 809, 167

\noindent Banzatti, A., Meyer, M.\ R., Manara, C.\ F., 
Pontoppidan, K.\ M., \& Testi, L.\ 2014, ApJ, 780, 26 

\noindent Bast, J.\ E., Brown, J.\ M., Herczeg, G.\ J., van Dishoeck, E.\ F.,
\& Pontoppidan, K.\ M.\ 2011, A\&A, 527, 119

\noindent Brown, L.\ R., Troutman, M., R., Gibb, E.\ L.\ 2013, ApJ,770, L14

\noindent Brown, J.\ M., Pontoppidan, K.\ M.\ van Dishoeck, E.\ F., Herczeg, G.\ J., Blake, G.\ A., Smette, A.\ 2013, ApJ, 770, 94

\noindent Bruderer, S., Harsono, D., \& van Dishoeck, E.\ F.\ 2015, A\&A, 575, 94

\noindent Carr, J.\ S.\ \& Najita, J.\ R.\ 2008, Science, 319, 1504

\noindent Carr, J.\ S.\ \& Najita, J.\ R.\ 2011, ApJ, 733, 102


\noindent Ciesla, F.\ J.\ \& Cuzzi, J.\ N.\ 2006, Icarus, 181, 178

\noindent Dauphas, M., \& Chaussidon, M.\ 2011, AREPS, 39, 351

\noindent Dello Russo, N., Kawakita, H., Vervack, R. J., Weaver, H. A. 2016, Icarus, 278, 301

\noindent Doppmann, G.\ W., Najita, J.\ R., Carr, J.\ S.\ 2008, ApJ, 685, 298

\noindent Edwards, S., Hartigan, P., Ghandour, L., \& Andrulis, C.\ 
1994, AJ, 108, 1056 

\noindent Eisner, J.\ A., Hillenbrand, L.\ A., White, R.\ J.,
Akeson, R.\ L., \& Sargent, A.\ I.\ 2005, ApJ, 623, 952 

\noindent Folha, D.\ F.\ M., \& Emerson, J.\ P.\ 2001, A\&A, 365, 90

\noindent Font, A.\ S., McCarthy, I.\ G., Johnstone, D., \& 
Ballantyne, D., R.\ 2004, ApJ, 607, 890 

\noindent Furlan, E., Hartmann, L., Calvet, N., et al.\ 2006, ApJS, 165, 568

\noindent Gibb, E.\ L., Van Brunt, K.\ A., Brittain, S.\ D., Rettig, T.\ W.\ 
2007, ApJ, 660, 1572

\noindent Gibb, E.\ L. \& Horne, D.\ 2013, ApJ, 776, L28

\noindent Greaves, J.\ S.\ 2004, MNRAS, 351, L99 

\noindent Hartigan, P., Edwards, S., \& Ghandour, L.\ 1995, 
ApJ, 452, 736

\noindent Isella, A., Carpenter, J.\ M., \& Sargent, A., I.\ 2009, 
ApJ, 701, 260

\noindent Johns-Krull, C. M., McLane, J. N., Prato, L. et al.\ 2016, ApJ, 826, 206

\noindent Kleine, T., Touboul, M., Bourdon, B., et al.\ 2009, Geo. Cosm. Acta, 73, 5150

\noindent Knez, C., Carr, J., Najita, J., Lacy, J., Richter, M., Bitner, M., 
Evans, N., II, van Dishoeck, E., Blake, G.\ 2007, AAS, 211, 5005

\noindent K\'osp\'al, \'A, Abraham, P., Acosta-Pulido, J. A., et al.\ 2012, ApJS, 201, 11

\noindent Lacy, J.\ H., Richter, M.\ J., Greathouse, T.\ K., 
Jaffe, D.\ T., Zhu, Q.\ 2002, PASP, 114, 153

\noindent Liu, M. C., Graham, J. R., Ghez, A. M., et al.\ 1996, ApJ, 461, 334

\noindent Loinard, L., Torres, R.~M., Mioduszewski, A.~J., 
\& Rodr{\'{\i}}guez, L.~F.\ 2008, \apjl, 675, L29 

\noindent Mamajek, E.~E.\ 2008, Astronomische Nachrichten, 329, 10 

\noindent Mandell, A.\ M., Bast, J., van Dishoeck, E.\ F., 
Blake, G.\ A., Salyk, C., Mumma, M.\ J., Villanueva, G.\ 2012, ApJ, 747, 92

\noindent McCabe, C., Ghez, A.\ M., Prato, L., 
Duch\^ene, G., Fisher, R.\ S., \& Telesco, C.\ 2006, 
ApJ, 636, 932  

\noindent Melo, C.\ H.\ F.\ 2003, A\&A, 410, 269

\noindent Muzerolle, J., Hartmann, L., \& Calvet, N.\ 1998, AJ, 
116, 2965 

\noindent Muzerolle, J., Hartmann, L., \& Calvet, N.\ 1998, AJ, 
116, 455 

\noindent Najita, J., Carr, J., S., \& Mathieu, R.\ D.\ 2003, ApJ, 589 931

\noindent Najita, J.\ R., Carr, J.\ S., Pontoppidan, K.\ M., Salyk, C., 
van Dishoeck, E.\ F., \& Blake, G.\ 2013, ApJ, 766, 134

\noindent Najita, J.\ R., \'Ad\'amkovics, M., \& Glassgold, A.\ E.\ 2011, ApJ, 743, 147

\noindent Najita, J.\ R., Andrews, S.\ M., \& Muzerolle, J.\ 2015, 
MNRAS, 450, 3559

\noindent Najita, J.\ R., \& \'Ad\'amkovics, M.\ 2017, ApJ, 847, 6 

\noindent Najita, J.\ R., \& Kenyon, S.\ J.\ 2014, MNRAS, 445, 3315

\noindent Nguyen, D.\ C., Brandeker, A., van Kerkwijk, M.\ H., 
\& Jayawardhana, R.\ 2012, ApJ, 745, 119 

\noindent \"Oberg, K. I., Guzm\'an, V. V., Furuya, K., et al.\ 2015, Nature, 520, 1980

\noindent Petrov, P.\ P., Gahm, G.\ F., Stempels, H.\ C., 
Walter, F.\ M., \& Artemenko, S.\ A., 2011, A\&A, 535, 6

\noindent Pontoppidan, K.~M., Salyk, C., Bergin, E.~A., 
et al.\ 2014, in Protostars and Planets VI, ed.\ H.\ Beuther, 
R.\ S.\ Klessen, C.\ P.\ Dullemond, \& T.\ Henning, 
(Tucson: University of Arizona), 363 

\noindent Pontoppidan, K.\ M., Salyk, C., Blake, G.\ A., Meijerink, R.,
Carr, J.\ S., \& Najita, J.\ 2010a, ApJ, 720, 887

\noindent Pontoppidan, K.\ M., Salyk, C., Blake, G.\ A., K\"aufl, H.\ U.\ 
2010b, ApJL, 722, L173

\noindent Pontoppidan, K.\ M., Blake, G.\ A., Smette, A.\ 2011, ApJ, 733, 84

\noindent Rothman, L.~S., Gordon, I.~E., Babikov, Y., et al.\ 2013, JQSRT, 130, 4 

\noindent Salyk, C., Pontoppidan, K.\ M., Blake, G.\ A., 
Lahuis, f., van Dishoeck, E.\ F., \& Evans, N.\ J.\ II 2008, 
ApJ, 676, L49

\noindent Salyk, C., Pontoppidan, K., Corder, S., Mu\~noz, D., 
Zhang, K., \& Blake, G.\ A.\ 2014, ApJ, 792, 68

\noindent Salyk, C., Pontoppidan, K.\ M., Blake, G.\ A., Najita, J.\ R., \& Carr, J.\ S.\
2011a, ApJ, 731, 130

\noindent Salyk, C., Blake, G.\ A., Boogert, A.\ C.\ A., \& Brown, J.\ M.\ 2011b, ApJ, 743, 112

\noindent Salyk, C., Lacy, J.\ H., Richter, M.\ J., Zhang, K., Blake, G.\ A., 
Pontoppidan, K.\ M.\  2015, ApJ, 810, L24

\noindent Schindhelm, E., France, K., Burgh, E.~B., et al.\ 2012, \apj, 746, 97 

\noindent Siess, L., Dufour, E., \& Forestini, M.\ 2000, 
A\&A, 358, 593

\noindent Simon, M., Pascucci, I., Edwards, S., et al.\ 2016, ApJ, 831, 169

\noindent Thi, W. F., van Dishoeck, E. F., Blake, G. A., et al.\ 2001, ApJ, 561, 1074

\noindent Wilking, B.\ A., Gagn\'e, M., \& Allen, L.\ E. 2008, in Handbook of Star Forming
Regions, Vol.\ 2, ed.\ B.\ Reipurth (San Francisco, CA: ASP), 351

\clearpage

\begin{figure}
\figurenum{1}
\epsscale{0.8}
\plotone{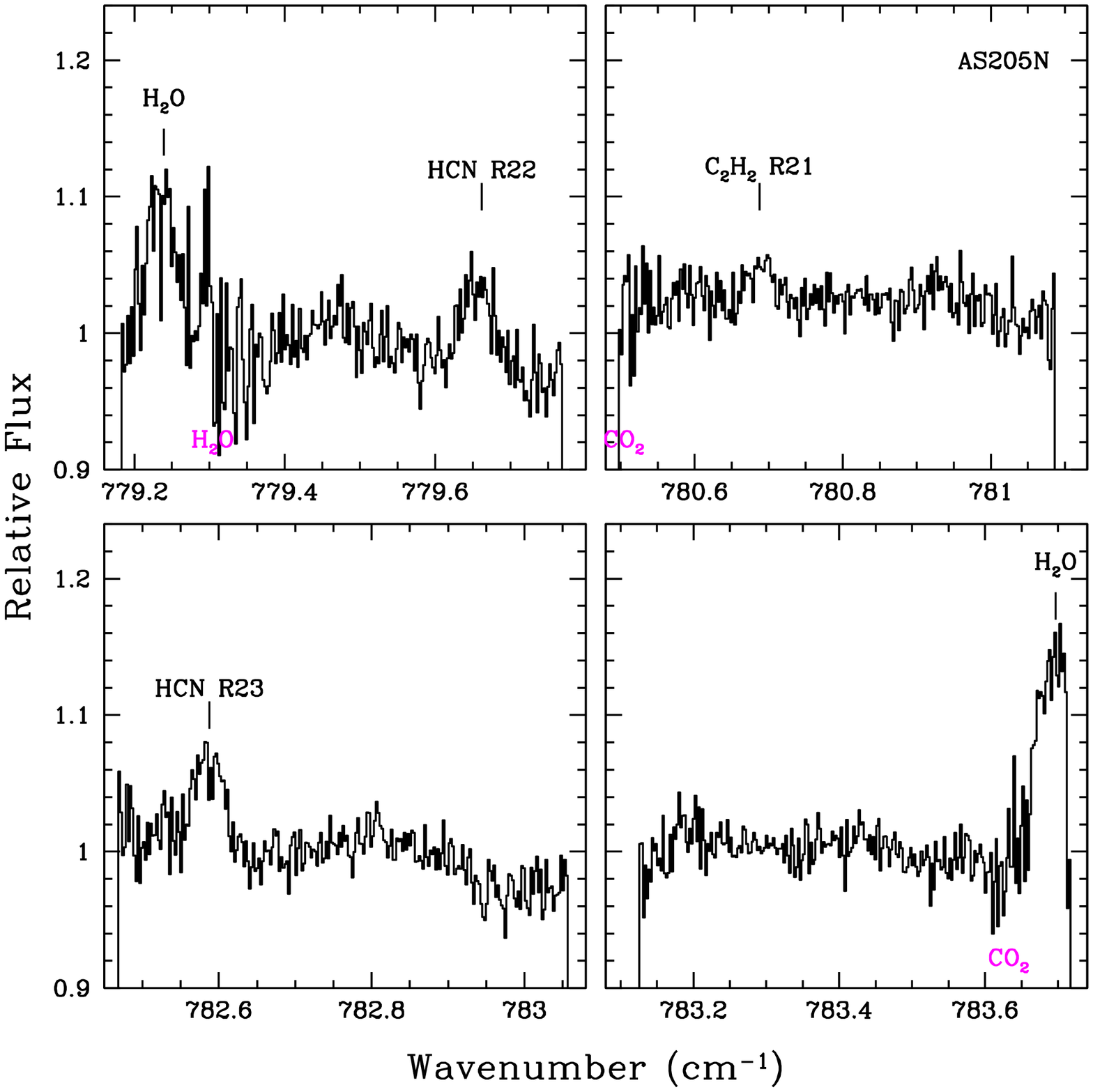}
\caption[]
{TEXES spectrum of \as205n\ in the 781\pcm\ setting showing 
the detection of HCN R22 and R23, \ctwohtwo\ R21, and water 
emission lines in the geocentric frame. 
Only orders containing detected features are shown. 
Emission features are labeled at the source velocity.
The position of telluric absorption features that cause 
increased noise are denoted below the spectrum 
(magenta labels). 
}
\end{figure}

\begin{figure}
\figurenum{2}
\plotone{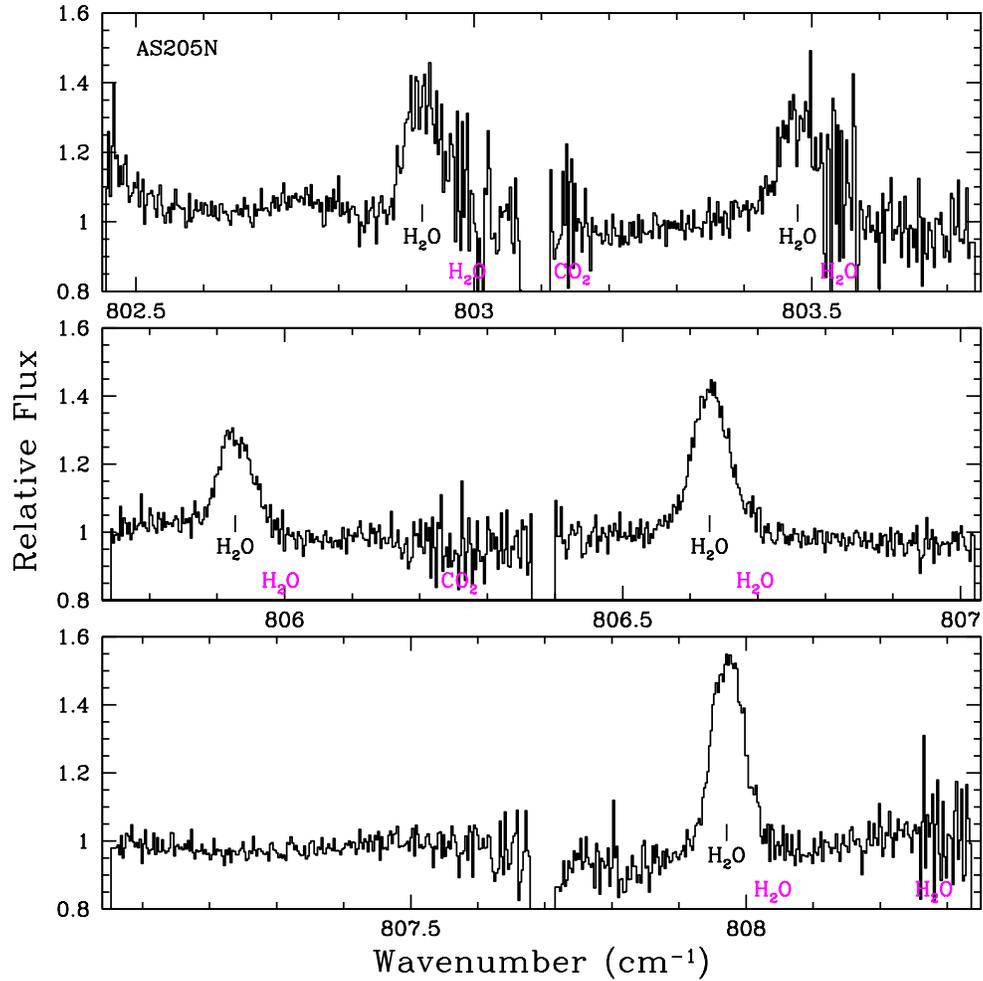}
\caption[]
{TEXES spectrum of \as205n\ in the 805\pcm\ setting showing 
the detection of bright water emission lines 
in the geocentric frame. 
The order near 807.5\pcm, which has no detected features, 
illustrates the typical signal-to-noise ratio in the absence of 
emission and strong telluric features.  
}
\end{figure}

\begin{figure}
\figurenum{3}
\epsscale{0.8}
\plotone{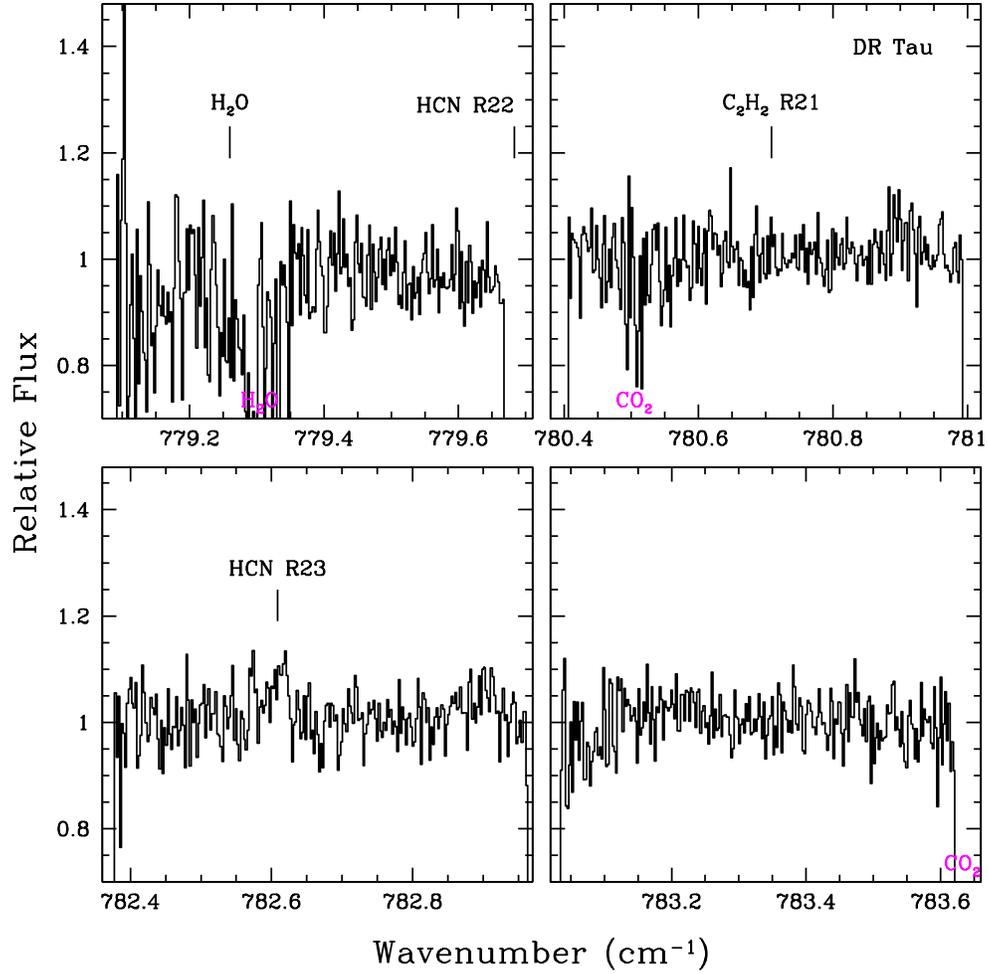}
\caption[]
{TEXES spectrum of DR Tau in the 781\pcm\ setting showing 
the detection of the HCN R23 line 
in the geocentric frame. 
The same orders shown in Figure 1 are plotted.
The HCN R22 line fell between two 
orders, and the \ctwohtwo\ R21 line was not detected. 
}
\end{figure}

\begin{figure}
\figurenum{4}
\epsscale{0.8}
\plotone{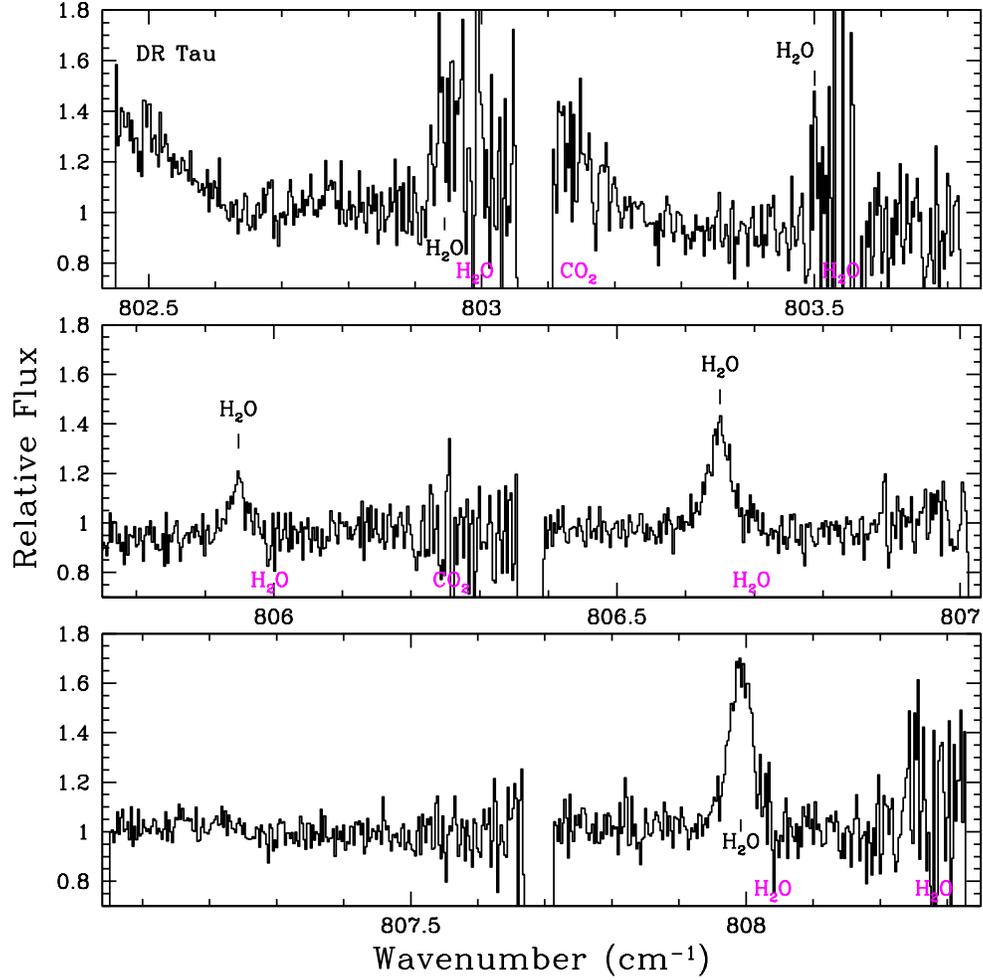}
\caption[]
{TEXES spectrum of DR Tau in the 805\pcm\ setting showing 
the detection of bright water emission lines 
in the geocentric frame. 
The same orders shown in Figure 2 are plotted.
}
\end{figure}

\begin{figure}
\figurenum{5}
\epsscale{0.5}
\plotone{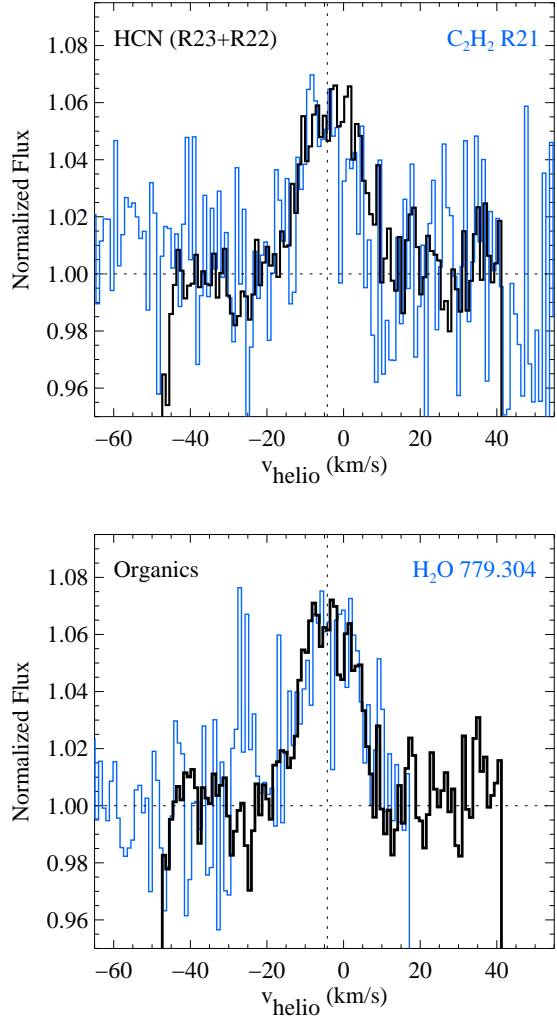}
\caption[]
{Comparison of organics and water line profiles 
in the spectrum of \as205n\ at the 781\pcm\ setting. 
In the top panel, 
the average HCN (thick black line) and 
\ctwohtwo\ R21 (thin blue line) profiles are compared,  
and in the bottom panel the average of all three organic lines  
(HCN R22, HCN R23, \ctwohtwo\ R21; thick black line) 
is compared with the 779\pcm\ water line (thin blue line). 
The vertical scale is the normalized flux of the average of 
the two HCN lines (top) and 
the three organic lines (bottom), 
with the comparison feature 
scaled to the same peak flux. 
The dotted vertical line marks 
the peak velocity of the strong \water\ lines 
in the 805\pcm\ setting.
In the bottom panel, the telluric water line, centered at 
-29.5\kms, may contribute residual water emission. 
}
\end{figure}

\begin{figure}
\figurenum{6}
\epsscale{0.5}
\plotone{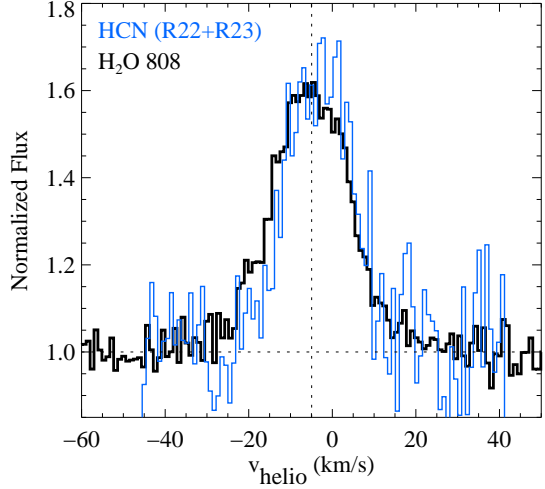}
\caption[]
{Comparison of the average HCN line profile (thin blue line) and 
the line profile of the 808\pcm\ water line (thick black line) 
for \as205n. 
The vertical scale is the normalized flux of the 
water line 
with the HCN 
scaled to the same peak flux. 
The vertical dashed line marks the 
centroid velocity of the three bright \water\ lines 
in the 805\pcm\ setting.  
}
\end{figure}

\begin{figure}
\figurenum{7}
\epsscale{0.5}
\plotone{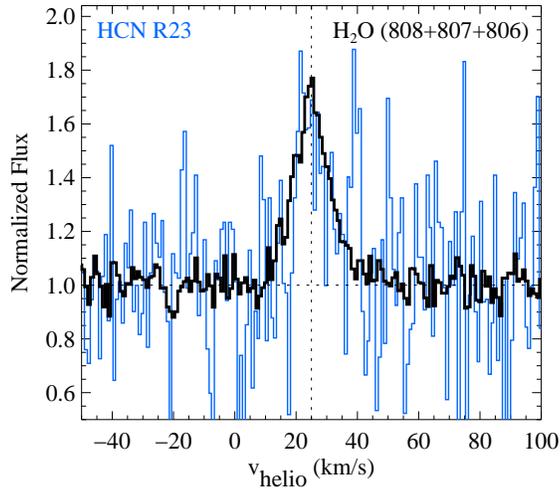}
\caption[]
{Comparison of the HCN R23 line profile (thin blue line) and 
the average profile of the three bright water lines 
observed in the 805\pcm\ setting (thick black line) 
in the DR Tau spectrum. 
The water emission is centered at 25\kms\ (vertical dashed line). 
}
\end{figure}

\begin{figure}
\figurenum{8}
\includegraphics[angle=270,scale=0.33]{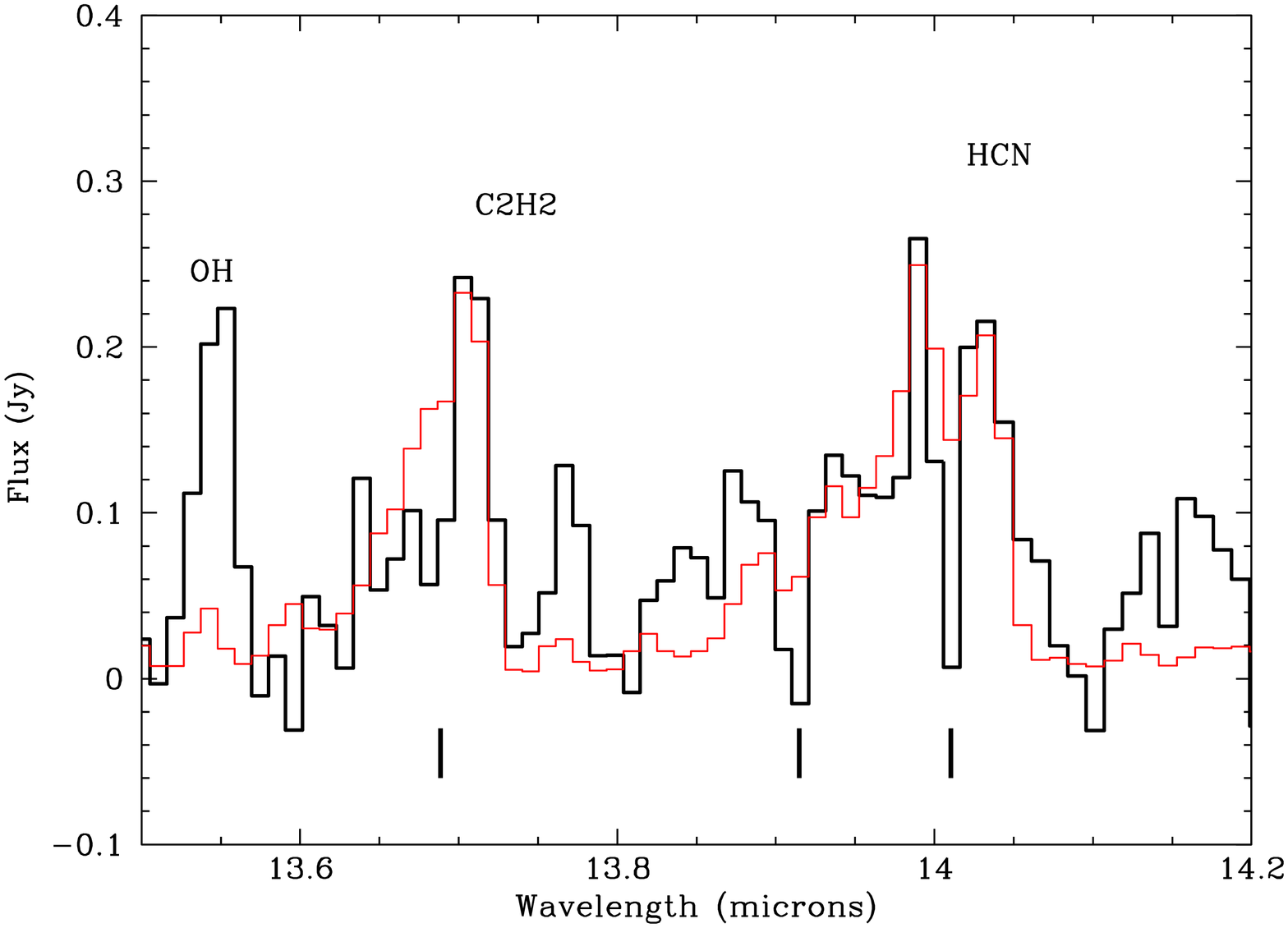}
\caption[]
{{\it Spitzer} IRS spectrum of HCN and \ctwohtwo\ emission
from \as205n\ (black line) compared with an LTE slab model (red line).
The water emission model (Fig.\ 18) has been subtracted from
the observed spectrum.
The model spectrum assumes 
HCN and \ctwohtwo\ column densities of
$N({\rm HCN}) = 4.2\times 10^{15}\psqcm$ and
$N(\ctwohtwo) = 1.5\times 10^{15}\psqcm$ and adopts
the same temperature (680\,K)
and emitting area as those for water.
Pixels affected by water emission features 
at the marked locations (vertical lines)
were excluded from the model fit. 
}
\end{figure}

\begin{figure}
\figurenum{9}
\includegraphics[angle=270,scale=0.33]{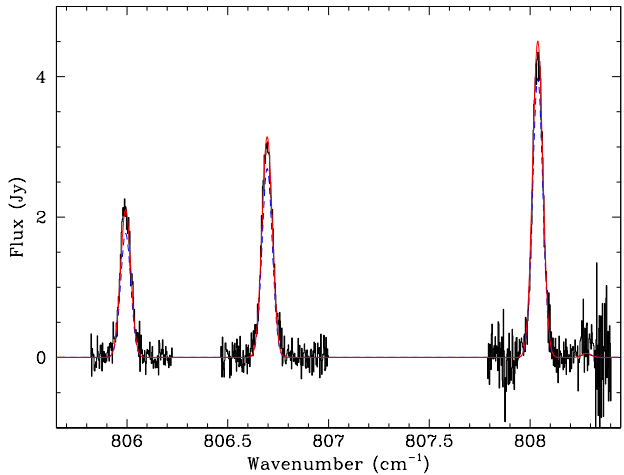}
\caption[]
{The three cleanest water emission lines in the
(continuum subtracted) TEXES spectrum of \as205n\
(black line), shown in the rest frame of the emission,
compared with the LTE emission model (dashed blue line)
used to fit the water
emission in the {\it Spitzer} IRS spectrum (Fig.\ 18)
and to a model with
$N(\water)$ increased to $1.6\times 10^{18}\psqcm$
to match the observed line fluxes (red line).
The model line emission has been
broadened by a Gaussian profile with a FWHM of 23\kms.
}
\end{figure}

\begin{figure}
\figurenum{10}
\includegraphics[angle=270,scale=0.33]{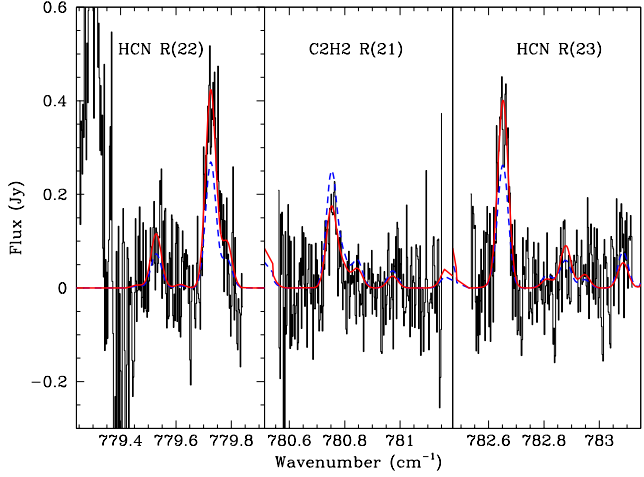}
\caption[]
{The HCN and \ctwohtwo\ lines observed in the continuum-subtracted
TEXES spectrum of \as205n, shown in the rest frame of the emission,
compared with the LTE emission model (dashed blue line) used to fit the
HCN and \ctwohtwo\ bands in the IRS spectrum (Fig.\ 8). The solid
red line shows a model with column densities adjusted to fit the
TEXES line fluxes, in which 
$N({\rm HCN}) = 6.8\times 10^{15}\psqcm$ and
$N(\ctwohtwo) = 1.0\times 10^{15}\psqcm$.
}
\end{figure}

\begin{figure}
\figurenum{11}
\includegraphics[angle=270,scale=0.33]{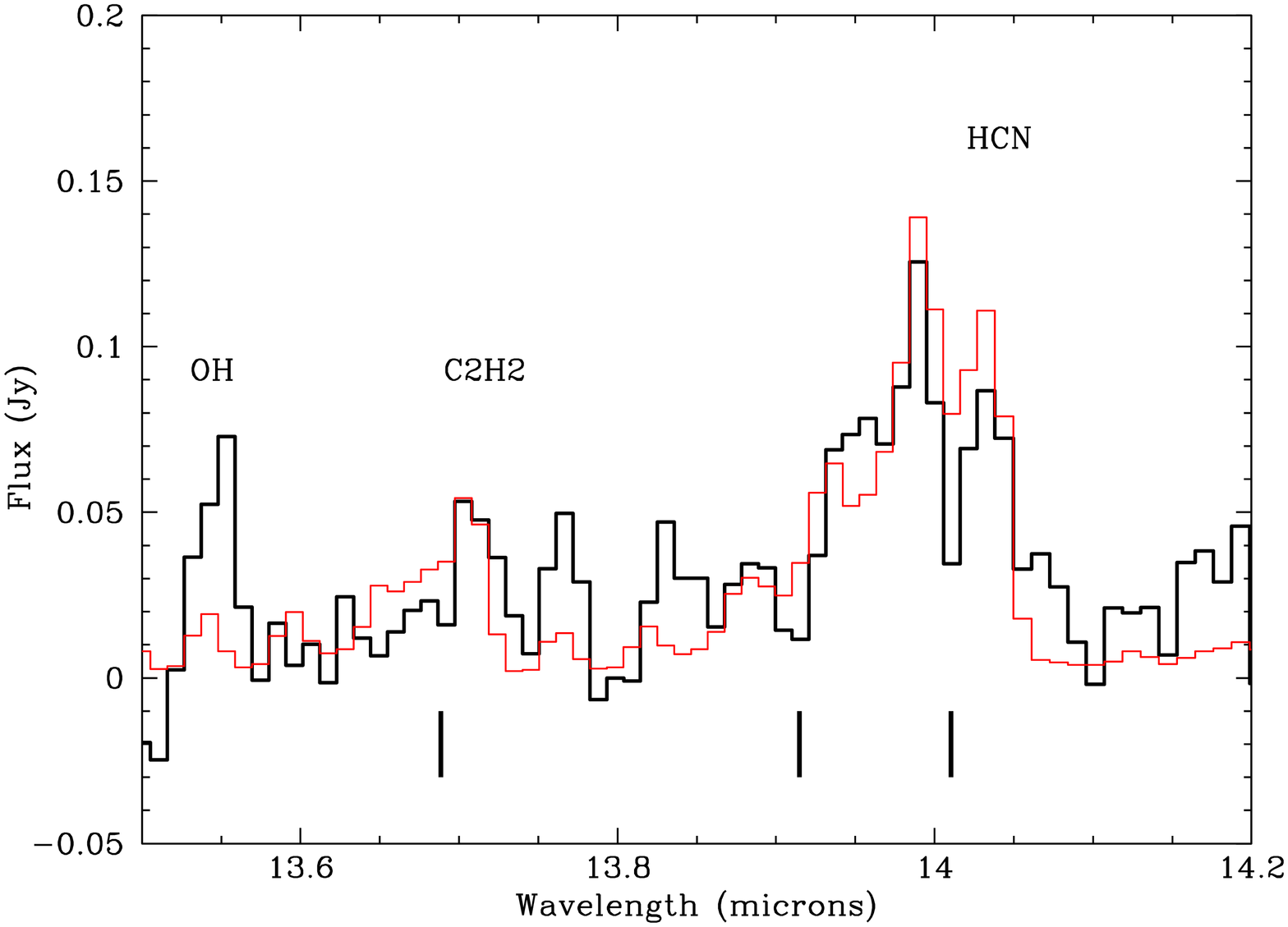}
\caption[]
{{\it Spitzer} IRS spectrum of HCN and \ctwohtwo\ emission from
DR Tau (black line) 
compared with an LTE slab model (red line).
The water model (Fig.\ 19) has been subtracted
from the observed spectrum.
The model spectrum uses HCN and \ctwohtwo\
column densities of
$N({\rm HCN}) = 5.7\times 10^{15}\psqcm$ and
$N(\ctwohtwo) = 6.8\times 10^{14}\psqcm$
and adopts the same temperature and emitting area as for
the water emission.
Pixels affected by water emission features 
at the marked locations (vertical lines)
were excluded from the model fit. 
}
\end{figure}

\begin{figure}
\figurenum{12}
\includegraphics[angle=270,scale=0.33]{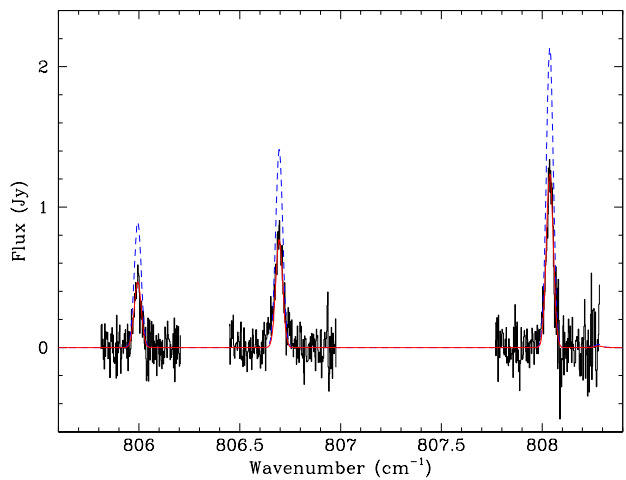}
\caption[]
{The three cleanest water emission lines in the
continuum-subtracted TEXES spectrum of
DR Tau (black line), 
shown in the rest frame of the emission.
The LTE model used to fit the \water\
emission in the {\it Spitzer} IRS spectrum of DR Tau (Fig.\ 19)
overpredicts the TEXES water emission (dashed blue line).
Lowering the \water\
column density to $4.6\times 10^{17}\psqcm$ produces a good
fit (red line). In the model spectra, the emission lines were
broadened by a Gaussian profile with a FWHM of 15\kms.
}
\end{figure}

\begin{figure}
\figurenum{13}
\includegraphics[angle=0,scale=0.45]{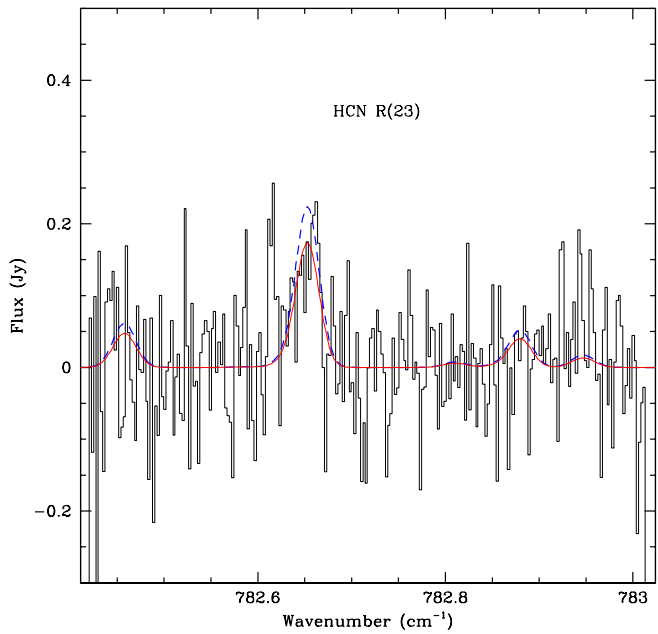}
%
\caption[]
{The HCN R(23) line observed in the continuum-subtracted
TEXES spectrum of DR Tau (black line), 
shown in the rest frame of the emission,
compared with the LTE model (dashed blue line) used to
fit the HCN band in the {\it Spitzer} IRS spectrum (Fig.\ 11)
and to a model (solid red line) with the column density decreased
to $N({\rm HCN}) = 4.4\times 10^{15}\psqcm$ in order 
to fit the line flux.
}

\end{figure}

\begin{figure}
\figurenum{14}
\epsscale{0.6}
\plotone{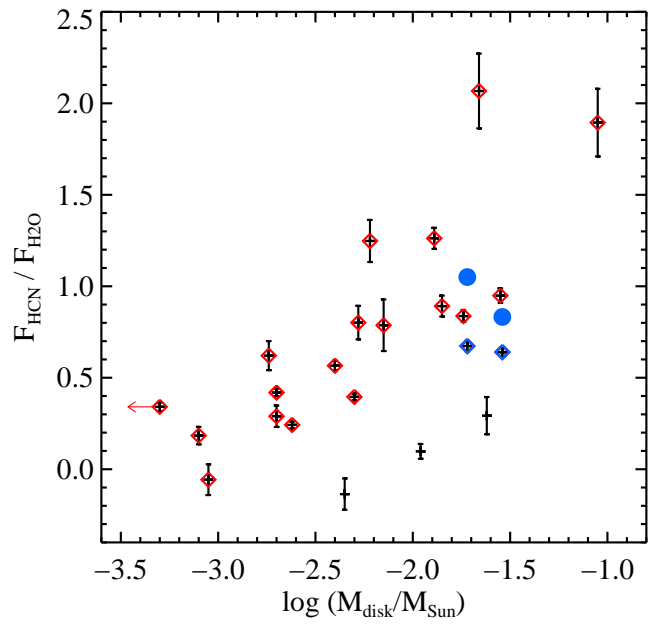}
\caption[]
{HCN/\water\ flux ratio of inner disks vs.\ disk mass 
from fits to low-resolution {\it Spitzer} IRS spectra 
(red diamonds; Najita et al.\ 2013) 
and the results obtained here 
for \as205n\ and DR Tau 
from their IRS (blue diamonds) and 
TEXES (blue circles) spectra. 
The latter assume that the change in the column density 
ratio between the two epochs reflects time variability 
in the flux ratio (see text for details). 
}
\end{figure}
\clearpage

\begin{figure}
\figurenum{15}
\epsscale{0.5}
\plotone{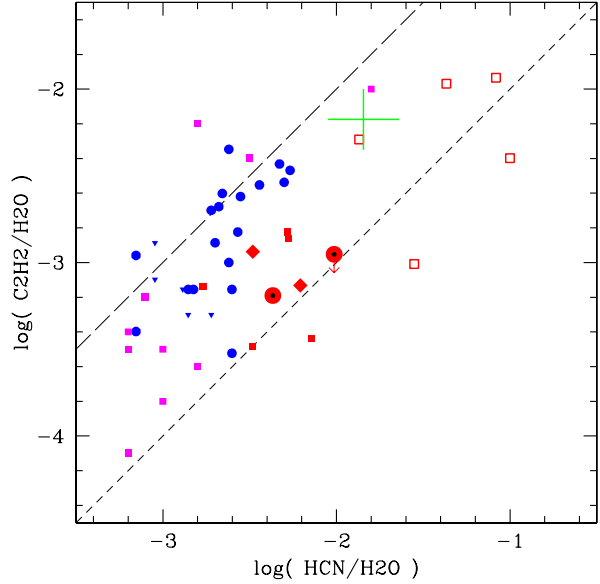}
%
\caption[]
{Column density ratios of HCN and \ctwohtwo\ relative to water for 
LTE slab fits to {\it Spitzer} IRS and TEXES spectra 
compared to cometary abundances. 
Model fits to IRS spectra that have equal emitting areas for 
HCN, \ctwohtwo, and water 
(red squares---Carr \& Najita 2011; 
magenta symbols---Salyk et al.\ 2011a) 
are closer to the 
abundances of comets (blue symbols; Dello Russo et al.\ 2016) 
than models that allow for different emitting areas of 
HCN and water (open red squares; Carr \& Najita 2011). 
The TEXES spectra indicate similar emitting areas for the HCN 
and water emission from \as205n\ and DR Tau 
(large red circles with black dots) and fall close to the 
fits to their IRS spectra assuming equal emitting areas 
(red diamonds). 
Several sources from Salyk et al.\ (2011a) fall outside the boundaries 
of the plot.
The range of abundances measured for hot cores are also 
shown (green cross; see Carr \& Najita 2011 for details).  
The dashed lines indicate constant ratios of \ctwohtwo/HCN = 0.1 and 1 
(short-dashed and long-dashed lines, respectively).
Blue inverted triangles indicate upper limits.
}
\end{figure}

\begin{figure}
\figurenum{16}
\epsscale{0.52}
\plotone{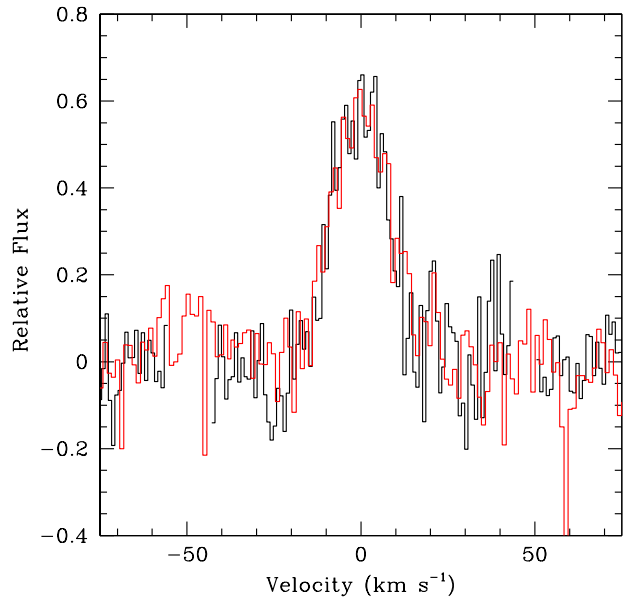}
%
\caption[]
{Comparison of the velocity profiles of the MIR and 
3\micron\ HCN emission lines in \as205n\ observed with 
TEXES and CRIRES, respectively. The MIR profile 
(black) is the average of the R(22) and R(23) transitions 
of the $\nu_2$ 
vibrational band. 
The 3\micron\ 
profile (red) is the average of the P(5), P(11), P(12) and P(13)
transitions of the 
$\nu_1$ band. 
The profiles have been shifted to center on zero velocity. 
The similarity of the line profiles differs dramatically from the 
non-LTE predictions of Bruderer et al.\ (2015). 
}
\end{figure}

\begin{figure}
\figurenum{17}
\includegraphics[angle=270,scale=0.33]{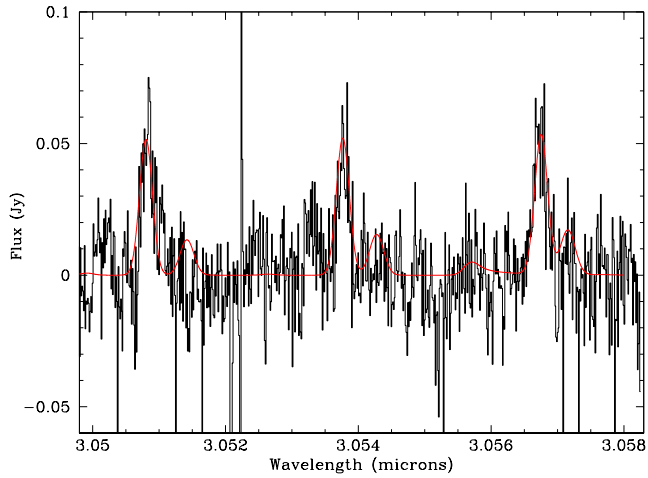}
\caption[]
{The CRIRES spectrum of HCN emission at 3\micron\ (black line) 
compared with an LTE model for the emission (red line) 
in the rest frame of the emission. 
The three plotted emission lines are the
P(11), P(12) and P(13) transitions of the 10$^0$0--00$^0$0 band.
The model is the same LTE model that fits the 14\micron\ Q branch in 
Fig.\ 8.  
}
\end{figure}

\begin{figure}
\figurenum{18}
\includegraphics[angle=270,scale=0.33]{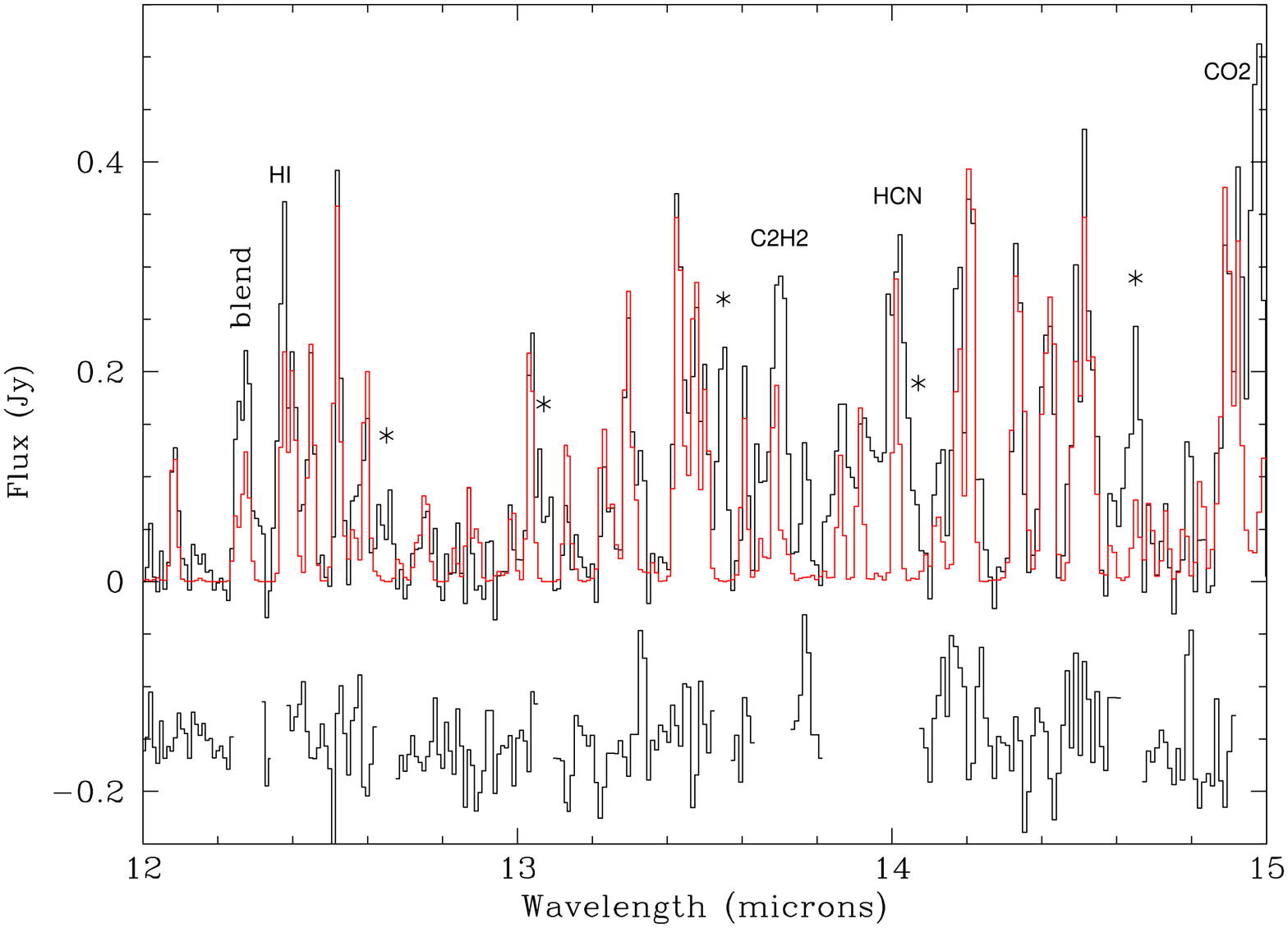}
\caption[]
{The {\it Spitzer} IRS spectrum of \as205n\ (black) compared to the best
fit water emission spectrum (red) calculated with an LTE slab
model. The adopted model parameters are
$T = 680$\,K,
$N(\water) = 1.3\times 10^{18}\psqcm$, and
$R_e = 1.90$\,AU
for the radius of the projected emitting area.
Emission from species other than water includes 
HCN at 14\micron,
\ctwohtwo\ at 13.7\micron,
OH (12.65, 13.07, 13.55, 14.07, 14.65\micron),
\cotwo\ at 15\micron, H I at 12.37\micron,
and a blend of features near 12.3\micron.
These features are marked in the spectrum,
with asterisks marking the position of OH features.
The difference between the water emission model and the observed
spectrum is also shown; spectral regions that are affected by
emission from other species are omitted.  
}
\end{figure}

\begin{figure}
\figurenum{19}
\includegraphics[angle=270,scale=0.33]{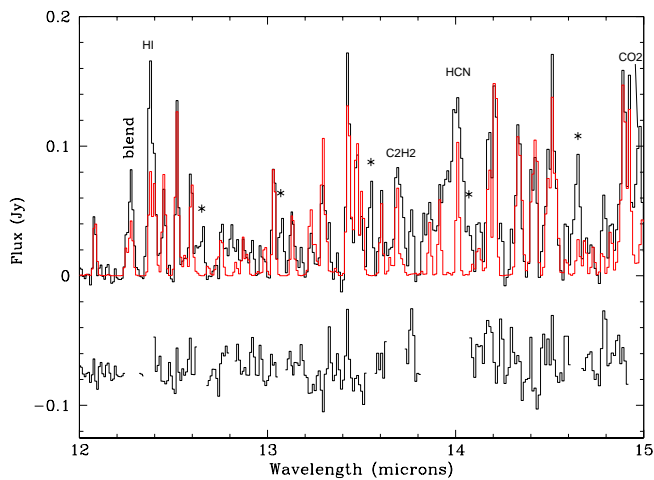}
\caption[]
{The {\it Spitzer} IRS spectrum of DR Tau (black) compared to the best-fit
water emission spectrum (red) calculated with a LTE slab model.
The adopted model parameters are
$T = 690$\,K,
$N(\water) = 9.3\times 10^{17}\psqcm$, and
$R_e = 1.33$\,AU
for the radius of the projected emitting area.
Emission from species other than water includes
HCN at 14\micron,
\ctwohtwo\ at 13.7\micron,
OH (12.65, 13.07, 13.55, 14.07, 14.65\micron),
\cotwo\ at 15\micron, H I at 12.37\micron,
and a blend of features near 12.3\micron.
These features are marked in the spectrum,
with asterisks marking the position of OH features.
The difference between the water emission model and the observed
spectrum is also shown; spectral regions that are affected by
emission from other species are omitted.  
} \end{figure}

\end{document}